\documentclass[twocolumn,amssymb, nobibnotes, aps, pra,nofootinbib,superscriptaddress]{revtex4-2}
\usepackage{amsmath}
\usepackage{braket}
\usepackage{graphicx}

\newcommand{\abs}[1]{\lvert #1 \rvert} 

\begin{document}

\title{Scalable High-Dimensional Multipartite  Entanglement with Trapped Ions}%

\newcommand{\iisc}{\affiliation{Department of Instrumentation and Applied Physics, Indian Institute of Science, Bangalore, India, 560012.}}

\author{Harsh Vardhan Upadhyay}
\email{harshv@iisc.ac.in}
\iisc

\author{Sanket Kumar Tripathy}
\iisc

\author{Ting Rei Tan}
\affiliation{School of Physics, University of Sydney, NSW 2006, Australia}

\author{Baladitya Suri}
\iisc

\author{Athreya Shankar}
\email{athreya@physics.iitm.ac.in}
\iisc

\date{\today}%

\begin{abstract}
    We propose a protocol for the preparation of generalized Greenberger-Horne-Zeilinger (GHZ) states of $N$ atoms each with $d=3$ or $4$ internal levels. We generalize the celebrated one-axis twisting (OAT) Hamiltonian for $N$ qubits to qudits by including OAT interactions of equal strengths between every pair of qudit levels, a protocol we call as balanced OAT (BOAT). Analogous to OAT for qubits, we find that starting from a product state of an \emph{arbitrary} number of atoms $N$, dynamics under BOAT leads to the formation of GHZ states for qutrits ($d=3$) and ququarts ($d=4$). While BOAT could potentially be realized on several platforms where all-to-all coupling is possible, here we propose specific implementations using trapped ion systems. We show that preparing these states with a fidelity above a threshold value rules out lower dimensional entanglement than that of the generalized GHZ states. For qutrits, we also propose a protocol to bound the fidelity that requires only global addressing of the ion crystal and single-shot readout of one of the levels. Our results open a path for the scalable generation and certification of high-dimensional multipartite entanglement on current atom-based quantum hardware.   
\end{abstract}
\maketitle

\section{Introduction}

The preparation and characterization of entangled states is a central ingredient in quantum physics and technologies. Paradigmatic studies of entanglement typically focus on bipartite systems with finite or infinite dimensional subsystems, or explore multipartite entanglement in ensembles of two-level systems. Recent efforts have aimed at exploring entanglement beyond these paradigms by attempting to prepare and certify genuine high-dimensional multipartite entanglement, involving $N\geq 3$ subsystems each with $d\geq 3$ dimensions. A prototypical state in this regard is the generalized Greenberger-Horne-Zeilinger (GHZ) state~\cite{bao2023verylarge,xing2023preparation}
\begin{equation}
    \ket{\Psi}_d^N = \frac{1}{\sqrt{d}}\sum_{\alpha=0}^{d-1} \ket{\alpha}^{\otimes N},
    \label{eqn:psi_gen_ghz}
\end{equation}
which describes an equal superposition of $N$ $d$-level qudits simultaneously being in one of the $d$ levels $\ket{\alpha}$. Such a high-dimensional multipartite entangled state can find potential applications in, e.g., quantum sensing~\cite{humphreys2013quantum,trenyi2024activation}, qudit-based quantum computing~\cite{chi_programmable_2022}, studies of violations of local realism~\cite{ryu2013greenberger,ryu2014multisetting,lawrence_rotational_2014}, and quantum communication~\cite{hillery_quantum_1999,qin2018efficient,bai2021verifiable,hu2021anovel,guo2022multiparty}. Furthermore, the remarkably simple structure of this state--- it is symmetric under permutation of particles as well as levels--- makes it an attractive target since these symmetries can be leveraged to both prepare the state as well as certify multipartite entanglement without the need to measure an extensive number of observables. For these reasons, this state has garnered significant attention in the photonics domain, where seminal experiments have prepared and certified the entanglement structure of few-particle, high-dimensional GHZ states~\cite{erhard2018experimental,imany2019highdimensional,bao2023verylarge}, while general schemes for arbitrary $N$ and $d$ have also been proposed~\cite{paesani2021scheme,bell2022protocol,xing2023preparation}. Inspired by these ideas, more recent works have realized the $3$-particle, $3$-dimensional GHZ state outside the photonics domain, by using qutrits encoded in artificial~\cite{lierta2022experimental} and real atoms~\cite{zhao2024dissipative}. In particular, the authors of Ref.~\cite{lierta2022experimental} presented an illuminating overview of the landscape of high-dimensional, multipartite entangled states and the status of their experimental realization. According to them, their experiment performed using $N=3$ transmon qutrits on an IBM device constituted the first preparation and certification of genuine high-dimensional multipartite entanglement outside of the photonics domain. However, routes to scale up the generalized GHZ states in either $N$ or $d$ outside the photonics domain  have remained relatively unexplored. An early study proposed a path to realize GHZ states of an arbitrary number of qutrits using Rydberg atoms traversing a microwave cavity~\cite{zou2004onestep}. However, this proposal remains unimplemented to the best of our knowledge, does not discuss entanglement certification, and does not straightforwardly generalize beyond qutrits. 

In this work, we propose a route to prepare and certify the entanglement structure of generalized GHZ states of an arbitrary number of qutrits ($d=3$) or ququarts ($d=4$) in quantum hardware with all-to-all coupling. Our contributions in this direction are fourfold: First, we introduce a multilevel generalization of the well-known one-axis twisting (OAT) Hamiltonian~\cite{kitagawa1993squeezed,pezze2018RMP}, which consists of OAT interactions of identical strengths between every pair of levels (Sec.~\ref{sec:prep}). We call this protocol balanced one-axis twisting (BOAT). We analyze the dynamics of initial product states under BOAT by extending the techniques employed by Agarwal and coworkers~\cite{agarwal1997atomic} in their analysis of qubits evolving under OAT. We find that BOAT can lead to the formation of generalized GHZ states for any number $N$ of qudits in dimensions $d=3$ and $d=4$. 

Second, we propose implementations of the qutrit and ququart versions of BOAT using trapped ion platforms (Sec.~\ref{sec:implementation}), which are now beginning to explore qudit-based quantum information processing~\cite{low2020practical,ringbauer2022auniversal,hrmo2023native}. We propose a qutrit realization using a string of ${}^{176}\rm{Lu}^+$ ions~\cite{kaewuam2019spectroscopy}. For the ququart case, we propose to leverage the optical-metastable-ground ($omg$) architecture available in certain ion species~\cite{allcock2021omg} to prepare exotic high-dimensional GHZ states spanning energy levels with very different characters. 

Third, we propose to certify the entanglement structure of the prepared state using a fidelity-based multipartite entanglement witness~\cite{malik2016multiphoton,fickler2014interface} (Sec.~\ref{sec:certification}). Previous works have shown that, for a system of $3$ qutrits, measuring $\mathcal{F}>2/3$ with respect to $\ket{\Psi}_3^3$ certifies $3$-dimensional entanglement for each of the 3 qutrits~\cite{erhard2018experimental,lierta2022experimental}. Here, we generalize these arguments and show that measuring $\mathcal{F}>\mathcal{F}_{\mathrm{th}}=(d-1)/d$ in a system of $N$ $d$-level qudits certifies at least $d$-dimensional entanglement in \emph{all} the $2^{N-1}-1$ possible bipartitions of the system. The observation is remarkable since only a small number of observables ($d^2$) are required to estimate $\mathcal{F}$, which, in turn, bounds the entanglement structure of an exponential number of bipartitions. Although the expression for $\mathcal{F}_{\mathrm{th}}$ has appeared previously~\cite{bao2023verylarge}, a discussion of its full implications for more than $3$ particles has not been published to the best of our knowledge. 

Fourth, we consider experimental scenarios where individual qudit control is not available, which is true for several trapped ion systems. In this case, a measurement of the fidelity $\mathcal{F}$ is typically hindered because the measurement of off-diagonal density matrix elements becomes challenging. Nevertheless, we introduce a protocol for qutrits that relies only on global control to measure the absolute values of the relevant off-diagonal matrix elements, which in turn can bound the fidelity $\mathcal{F}$ (Sec.~\ref{sec:fid_estimation}). Therefore, experimentally determining a lower bound $>\mathcal{F}_{\rm th}$ is sufficient to certify high-dimensional multipartite entanglement, without the need for individual qudit control. Our protocol takes inspiration from recent works that have demonstrated effective time-reversal of Hamiltonian dynamics, leading to the measurement of so-called multiple quantum coherence spectra~\cite{gärttner2017measuring,garttner2018relating,lewisswan2019unifying,colombo2022timereversal}.

\section{Preparation of generalized GHZ states}

\label{sec:prep}

In this section, we first provide a brief background on preparing GHZ states of qubits with conventional one-axis twisting (OAT). Subsequently, we introduce balanced one-axis twisting (BOAT) for qudits, study dynamics under BOAT for initial product states, and demonstrate that it can lead to $N$-particle $d$-dimensional GHZ states for any $N$ and for $d=3,4$. Finally, we discuss the form of the GHZ states produced by this protocol.

\subsection{Background: One-axis twisting with qubits}

Here, we briefly summarize the techniques and key findings of Ref.~\cite{agarwal1997atomic} that are relevant to our present work. The system studied in Ref.~\cite{agarwal1997atomic} is a collection of $N$ atoms, modeled as qubits, collectively coupled to a cavity mode. The qubits are assumed to be identical and their transition frequency is far detuned from the cavity resonance. In this regime, the cavity mode can be eliminated, resulting in a collective dispersive shift on the atoms given by the Hamiltonian 
\begin{align}
    H = \chi S_+S_- = \chi (S^2 - S_z^2 + S_z).
    \label{eqn:ham_qubit}
\end{align}
Here, we have introduced collective spin operators $S_\nu = \sum_{j=1}^N \sigma_\nu^j/2, \; \nu=x,y,z$, with $\sigma_\nu^j$ the usual Pauli matrices for qubit $j$. The collective raising and lowering operators are given by $S_{\pm} = S_x \pm i S_y = \sum_{j=1}^N \sigma_\pm^j$. The total angular momentum  is given by $\mathbf{S} = \sum_{\nu\in \{x,y,z\}} S_\nu \boldsymbol{\hat{\nu}}$, using which we define $S^2 = \mathbf{S}\cdot \mathbf{S}$. The nonlinear term $S_z^2$ appearing in Eq.~(\ref{eqn:ham_qubit}) is responsible for the well-known one-axis twisting (OAT) dynamics which can lead to the formation of entangled spin states.

Starting with all qubits in the lower level $\ket{0}$, a global rotation pulse prepares the qubits in an initial coherent spin state with equal population in both levels given by 
\begin{align}
\ket{\psi(0)} \equiv \ket{ \phi'}=&\frac{1}{2^{N/2}}\left[\ket{0} + e^{i\phi'}\ket{1}\right]^{\otimes N}.
\end{align}
This state can be written in terms of the Dicke states as
\begin{align}
\ket{\psi(0)}=&\sum_{l=0}^N C_{N,l} e^{i l \phi'} \ket{N,l},
\end{align}
where
\begin{align}
    C_{N,l} = \frac{1}{2^{N/2}} \sqrt{\frac{N !}{(N-l) ! l !}}.
\end{align}
Here, $l$ denotes the number of atoms in $\ket{1}$, and the basis states $\{\ket{N,l}\}$ are the particle-permutation symmetric Dicke states, i.e., the eigenstates of $S_z$ with eigenvalues $l-N/2$. These basis states are also simultaneous eigenstates of $H$ with eigenvalues given by 
\begin{align}
\lambda_{l} =\chi((N+1)l - l^2). 
\end{align}
Hence, absorbing the terms linear in $l$ into the definition of the azimuthal phase, the state at time $\tau$ can be written as a function of $\chi\tau$ as 
\begin{equation}
\ket{\psi(\chi\tau)} =\sum_{l=0}^N \mathcal C_{N,l}  e^{ i l \phi} e^{i \chi\tau l^2} \ket{N,l},
\label{eqn:psi_tau_qubit}
\end{equation}
where $\phi = \phi' - \chi\tau \left(N+1\right)$. A constant term in the eigenvalue arising from the $S^2$ term is ignored as it will just give a global phase.

The key insight of Ref.~\cite{agarwal1997atomic} is that, at the special times $\chi\tau=2\pi/m, m=1,2,\ldots$, the term $e^{i\chi\tau l^2}$ is periodic in $l$ with a period $m$, and hence can be expanded in a Fourier series as 
\begin{align} 
e^{2\pi i \frac{l^2}{m}} = \sum_{q = 0}^{m-1}  f_{q} e^{2\pi i \frac {q l}{m}}, 
\end{align}
where $f_q$ are Fourier coefficients. Substituting this Fourier series in Eq.~(\ref{eqn:psi_tau_qubit}) and reordering the summation over $l$ and $q$ leads to the central observation that $\ket{\psi(2\pi/m)}$ can be expressed as a superposition of a discrete number of coherent spin states:
\begin{align}
    \ket{\psi(2\pi/m)} = \sum_{q=0}^m f_q \ket{\phi_q},
\end{align}
where $\phi_q = \phi + 2\pi q/m$. 

In particular, when choosing $m=4$, only two Fourier coefficients are non-zero and, moreover, they have equal magnitude, resulting in
\begin{align}
\begin{split}
    \ket{\psi(\pi/2)} &= \frac{1}{\sqrt{2}}\left| \phi' - (N-1)\frac{\pi}{2}\right\rangle\\ &- 
    \frac{i}{\sqrt{2}}\left|  \phi' - (N+1)\frac{\pi}{2}\right\rangle.
\end{split}
\end{align}
Hence, the initial coherent state evolves into a superposition of two orthogonal coherent states and is thus a GHZ state. Global single-qubit rotations can be applied to this state to align the two orthogonal states along the $z$ direction, resulting in the more familiar form $\ket{\psi}_{\rm GHZ} = \left(\ket{0}^{\otimes N} + \ket{1}^{\otimes N} \right)/\sqrt{2}$. 

In the following sections, we take inspiration from this analysis and generalize the initial state, Hamiltonian and the techniques employed in order to explore the preparation of high-dimensional multipartite GHZ states.

\subsection{Generalizing to qudits: Balanced one-axis twisting}

We now consider a generalization of the OAT term $S_z^2$ in  Hamiltonian~(\ref{eqn:ham_qubit}) to a system of $N$ interacting qudits, each with $d$ levels, $\alpha=0,\ldots,d-1$. For every pair of levels $\alpha,\beta$, we denote the population difference operator by $\sigma_{\beta\alpha,z}^j = \ket{\beta}_j\bra{\beta}_j-\ket{\alpha}_j\bra{\alpha}_j$. We can then introduce a collective population difference operator $S_{\beta\alpha,z} = \sum_j \sigma_{\beta\alpha,z}^j/2$. We consider an interaction between the qudits described by the Hamiltonian 
\begin{align} 
    H = -\chi' \sum_{\alpha = 0}^{d-2} \sum_{\beta > \alpha}^{d-1} S_{\beta\alpha,z}^2.
    \label{eqn:ham_qudit}
 \end{align}
This Hamiltonian generalizes the OAT term $S_z^2$ for qubits to $d$-level systems by introducing a separate OAT-like term for every pair of levels. Furthermore, these terms appear with equal rates in the Hamiltonian, and hence we call this interaction as balanced OAT (BOAT). 

In the case of qubits, the initial state leading to the formation of a GHZ state was a coherent spin state with equal superposition of the two levels. Guided by this result and the identical interaction strength between all pairs of levels in Eq.~(\ref{eqn:ham_qudit}), we consider an initial state of the form 
\begin{align}
    \ket{\psi(0)} = \ket{\boldsymbol{\phi}'},
    \label{eqn:psi0_qudit}
\end{align}
where $\boldsymbol{\phi}' =\{\phi_1',\ldots,\phi_{d-1}'\}$ is a $d-1$ dimensional vector of phases that specifies an $N$ qudit product state as
\begin{align}
    \ket{\boldsymbol{\phi}'} = \left[\frac{1}{\sqrt{d}}\ket{0} + \frac{1}{\sqrt{d}}\sum_{\alpha=1}^{d-1} e^{i\phi_{\alpha}'} \ket{\alpha}\right]^{\otimes N}.
\end{align}

The Hamiltonian~(\ref{eqn:ham_qudit}) and initial state~(\ref{eqn:psi0_qudit}) are symmetric under the permutation of qudit indices. Hence, the dynamics is constrained within the space of particle permutation symmetric basis states, defined by 
\begin{align}
       \ket{N,\boldsymbol{l}} =\mathcal{S}(\ket{0}^{\otimes N-\sum_\alpha l_\alpha},\ket{1}^{\otimes l_1},\ldots,\ket{d-1}^{\otimes l_{d-1}}).
\end{align}
Here, $\boldsymbol{l} = \{l_1,\ldots,l_{d-1}\}$ is a $d-1$-dimensional vector describing the number of particles in levels $\ket{\alpha},\alpha=1,\ldots,d-1$ and the population in $\ket{0}$ is set by number conservation. The symbol $\mathcal{S}$ denotes the symmetrizer, denoting the operation of performing the sum over all permutations of particle indices for a given $\boldsymbol{l}$ and normalizing the resulting superposition state. Furthermore, the $\ket{N,\boldsymbol{l}}$ bases are eigenstates of the Hamiltonian~(\ref{eqn:ham_qudit}) with eigenvalues given by 
\begin{align}
    &\lambda_{\boldsymbol{l}} =  -\chi(g(\boldsymbol{l}) -  h(\boldsymbol{l})), \nonumber\\
    &g(\boldsymbol{l}) = \sum_{\alpha=1}^{d-1}\sum_{\beta=\alpha}^{d-1} l_\alpha l_\beta, \;
    h(\boldsymbol{l}) = \sum_{\alpha=1}^{d-1}N l_\alpha,
\end{align}
where $\chi =  (d/2)\chi'$. As in the qubit case, the dynamics can now be solved by expressing the initial state in the $\ket{N,\boldsymbol{l}}$ basis as 
\begin{align} 
\ket{\psi(0)} = \sum_{\substack{\boldsymbol{l} \\ l_\alpha\geq 0;\;\sum_\alpha l_\alpha \leq N}}  C_{N,\boldsymbol{l}} & e^{i \boldsymbol{\phi}' \cdot \boldsymbol{l}} \ket{N,\boldsymbol{l}},
\end{align}
where the probability amplitude is determined by a multinomial coefficient as 
\begin{align}
    C_{N,\boldsymbol{l}} = \frac{1}{d^{N/2}} \sqrt{\frac{N!}{\left(N-\sum_{\alpha=1}^{d-1} l_\alpha\right)!\left[ \Pi_{\alpha=1}^{d-1} (l_\alpha !)\right]}}. 
\end{align}
The time-evolved state is then given by 
\begin{align}
    \ket{\psi(\tau)} = \sum_{\substack{\boldsymbol{l} \\ l_\alpha\geq 0; \;\sum_\alpha l_\alpha \leq N}} C_{N,\boldsymbol{l}} e^{i\boldsymbol{\phi}(\tau)\cdot \boldsymbol{l}} e^{i\chi\tau g(\boldsymbol{l})}\ket{N,\boldsymbol{l}},
    \label{eqn:psi_tau}
\end{align}
where the terms linear in $\boldsymbol{l}$ arising from $h(\boldsymbol{l})$ have been absorbed into the phases $\boldsymbol{\phi}(\tau)$, whose elements are given by  $\phi_\alpha(\tau)=\phi_\alpha'- \chi\tau N$.

Analogous to the qubit case, we study the time evolved state at times $\chi\tau=2\pi/m$ for $m=1,2,\ldots$. At such times, the function $e^{2\pi i g(\boldsymbol{l})/m}$ is periodic in all the variables $l_\alpha$ with period $m$. Hence, it can be expressed in a $(d-1)$-dimensional Fourier series as 
\begin{align}
    e^{2\pi i \frac{g(\boldsymbol{l})}{m}} = \sum_{q_1,\ldots,q_{d-1}=0}^{m-1} f_{\boldsymbol{q}} e^{2\pi i \frac{\boldsymbol{q}\cdot\boldsymbol{l}}{m}},
\end{align}
where $\boldsymbol{q}=\{q_1,\ldots,q_{d-1}\}$ is the  vector of indices in Fourier space. The Fourier series representation once again makes the argument of the exponential terms linear in $\boldsymbol{l}$. Hence, substituting this expansion in Eq.~(\ref{eqn:psi_tau}) and interchanging the order of summation over $\boldsymbol{l}$ and $\boldsymbol{q}$ makes the time-evolved state amenable to an interpretation in terms of superpositions of a discrete number of qudit coherent states as
\begin{align}
    \ket{\psi(\tau)} = \sum_{\boldsymbol{q}} f_{\boldsymbol{q}} \ket{\boldsymbol{\phi}_{\boldsymbol{q}}},
    \label{eqn:fourier_qudit}
\end{align}
where $\boldsymbol{\phi}_{\boldsymbol{q}} = \boldsymbol{\phi} + 2\pi \boldsymbol{q}/m$.

In order for the state at time $\chi\tau=2\pi/m$ for some $m$ to qualify as a GHZ state, a sufficient condition is that the following two properties are satisfied: 
\begin{enumerate}
    \item Only $d$ of the Fourier coefficients $f_{\boldsymbol{q}}$ are non-zero and have equal magnitude, and
    \item The corresponding coherent states $\ket{\boldsymbol{\phi}_{\boldsymbol{q}}}$ are mutually orthogonal.
\end{enumerate}
The two conditions ensure that the state is a superposition of $d$ orthogonal coherent (product) states, which can then be rotated into the GHZ state~(\ref{eqn:psi_gen_ghz}) using only global single-particle unitary operations. 

We note that the Fourier series representation~(\ref{eqn:fourier_qudit}) is independent of the number of particles $N$ and only depends on $d$ and $m$. Hence, meeting the above sufficiency condition for a given $d$ and $m$ would imply that a $d$-dimensional GHZ state is formed at time $\chi\tau=2\pi/m$ for \emph{any} particle number $N$. 

\begin{table}[!htb]
  \begin{math}
  \label{tab:table1}
    \begin{tabular}{|l| c c c c r|} 
      \hline
      \textbf{$K_{m,d}$} & $m = 2$ & $m = 3$ & $m = 4$ & $m = 5$ & $m = 6$\\
      \hline
      $d = 2$ & 1 & 3 & \textbf{2} & 5 & 6/2\\
      $d = 3$ & $ 2^2 $ & \textbf{3} & $  4^2  $ & $5^2$ & $6^2/3$ \\
      $d = 4$ & \textbf{4} & $  3^3  $ & $4^2$ & $  5^3  $ & $6^3/2$\\
      $d = 5$ & $2^4$ & $3^4$ & $4^4$ & $5^3$ & $6^4$\\
      $d = 6$ & $2^4$ & $3^4$ & $4^5/2$ & $5^5$ & $6^4$\\
      $d = 7$ & $  2^6  $ & $3^6$ & $  4^6  $ & $5^6$ & $  6^6  $\\
      \hline
    \end{tabular}
    \end{math}
    \caption{$K_{m,d}$: Number of non-zero Fourier coefficients for a $d$-dimensional system at time $\tau=2\pi/m$.}
\label{tab:fourier_nonzero}
\end{table}

In order to check for criterion 1 above, we numerically determine the number of non-zero Fourier coefficients for different $d$ and $m$. The results are summarized in Table~\ref{tab:fourier_nonzero}. Empirically, we find that the number of non-zero Fourier coefficients is given by 
\begin{align}
    K_{m,d} = \frac{m^{d-1}}{\text{gcd}(m,d)},
\end{align}
where $\text{gcd}(m,d)$ is the greatest common divisor of $m$ and $d$. The three entries highlighted in bold in Table~\ref{tab:fourier_nonzero} denote the combinations of $d$ and $m$ where $K_{m,d} = d$ is satisfied. These entries indicate that, for $d=2,3$ and $4$, the time-evolved state at $\chi\tau=2\pi/m$ with $m=4,3$ and $2$ respectively can be represented as a superposition of exactly $d$ distinct coherent states. We further analyze these three cases and verify that the non-zero Fourier coefficients all have equal magnitudes and that the $d$ coherent states contributing to the superposition are mutually orthogonal. Hence, these three cases satisfy both criteria 1 and 2, and thus BOAT can lead to the formation of $N$ particle GHZ states in $d=2,3,4$. Curiously, for $d\geq 5$, the condition $K_{m,d}=d$ cannot be satisfied, and we find that dynamics under BOAT does not generate GHZ states. This feature of the $N$-qudit BOAT protocol is analogous to that reported in Ref.~\cite{ringbauer2022auniversal}, where a different type of qudit gate was found to lead to a maximally entangled state of two qudits for $d=3,4$ but not for $d=5$.
 
\subsubsection{GHZ states with qutrits ($d=3$)}

Here, the initial state is given by 
\begin{align} \begin{split} 
|\psi\rangle = |\phi_1',\phi_2'\rangle \equiv \left (\dfrac {1}{\sqrt{3}} \left(|0\rangle + e^{i\phi_1'} |1\rangle + e^{i\phi_2'} |2\rangle \right) \right)^{\otimes N}.
 \end{split}
 \label{eqn:qutrit_init}
 \end{align}
After evolving under BOAT for time $\chi\tau = 2\pi/m$, the time-evolved state can in general be represented as a superposition of $m^2$ coherent states. For $m=3$, we find that only $3$ of the $9$ possible Fourier coefficients are nonzero:
\begin{align} \begin{split} 
     f_{1,2} = f_{2,1} = e^{-i2\pi/3}f_{0,0} = \frac{1}{\sqrt{3}},
 \end{split}\end{align}
The subscripts in $f_{q_1,q_2}$ denote the elements of the $2$-dimensional vector of Fourier indices $\boldsymbol{q}$. Therefore, the state at time $\chi\tau = 2\pi/3$ is given by 

\begin{align}
        \begin{split}
            \lvert\psi(2\pi/3)\rangle &= \frac{1}{\sqrt{3}} e^{i2\pi/3}\left|\phi_1' - 2\pi \frac{N}{3},\phi_2' - 2\pi \frac{N}{3}\right>   \\ & + \frac{1}{\sqrt{3}}   \left|\phi_1'-2\pi\frac{N-1}{3},\phi_2'-2\pi\frac{N-2}{3}\right>   \\ &  + \frac{1}{\sqrt{3}}  \left|\phi_1'-2\pi\frac{N-2}{3},\phi_2'-2\pi\frac{N-1}{3}\right> .
        \end{split}
        \label{eqn:psi_3d_ghz}
\end{align}
As these $3$ states are orthogonal to each other, $\ket{\psi(2\pi/3)}$ is an $N$-particle $3$-dimensional GHZ state. 

A well-known feature of conventional OAT is that the orientation of the $2$ coherent spin states constituting the GHZ state depends on the parity of $N$. Interestingly, this feature generalizes to the $d=3$ case, where we find that the orientation of the $3$ coherent spin states in Eq.~(\ref{eqn:psi_3d_ghz}) is different for $N=3k,3k+1,3k+2$. In particular, we find that for $N=3k$, one of the coherent states constituting the GHZ is the initial state itself. On the other hand, for $N=3k+1, N=  3k+2$, one of the coherent states is the initial state shifted in both its phases by $-2\pi/3$ and $2\pi/3$ respectively.

\subsubsection{GHZ states with ququarts ($d=4$)}

Here, the initial state is:
\begin{align} \begin{split} 
|\psi\rangle =& |\phi_1',\phi_2',\phi_3'\rangle \\ \equiv& \left (\dfrac {1}{2} \left(|0\rangle + e^{i\phi_1'} |1\rangle + e^{i\phi_2'} |2\rangle + e^{i\phi_3'} |3\rangle \right) \right)^{\otimes N}.
 \end{split}\end{align}
At time $\chi\tau=\pi$, evolution under BOAT leads to the formation of an $N$-particle $4$-dimensional GHZ state given by 
\begin{align} \begin{split}
    |\psi(\pi)\rangle =  &\frac{1}{2}|\phi'_1 - N\pi,\phi'_2- (N-1)\pi,\phi'_3-(N-1)\pi\rangle \\ + &\frac{1}{2}|\phi'_1-(N-1)\pi,\phi'_2- N\pi,\phi'_3-(N-1)\pi\rangle \\ + &\frac{1}{2}|\phi'_1-(N-1)\pi,\phi'_2-(N-1)\pi,\phi'_3- N\pi\rangle \\ - &\frac{1}{2}|\phi'_1- N\pi,\phi'_2- N\pi,\phi'_3- N\pi\rangle.
    \end{split}\end{align}
The orientation of the coherent states constituting the GHZ state depends once again only on the parity of $N$, similar to conventional OAT; for even $N$ (odd $N$), one of the coherent states is the initial state itself (initial state displaced by angle $\pi$).

\section{Implementation with trapped ions}
\label{sec:implementation}

In this section, we first discuss how BOAT can be realized using serial entangling operations in all-to-all coupled systems. This observation paves the way to design specific experimental schemes using trapped ions, which we discuss subsequently.

\subsection{Serial one-axis twisting using two-level entangling operations}
\label{sec:serial_oat}
An attractive feature of BOAT is that the different terms in the Hamiltonian~(\ref{eqn:ham_qudit}) commute with each other. This opens the possibility to realize BOAT by sequentially implementing conventional OAT interactions between different pairs of levels. Moreover, OAT interactions between any pair of levels can be realized using OAT between a \emph{fixed} pair of levels by sandwiching the latter operation between appropriate global single-qudit gates that swap different levels. For instance, 
\begin{equation}
    S_{\gamma\delta,z}^2 = U_{\gamma\alpha}U_{\delta\beta}S_{\alpha\beta,z}^2 U_{\gamma\alpha}^\dag U_{\delta\beta}^\dag, 
\end{equation}
where the single-qudit swapping unitaries take the form 
\begin{equation}
    U_{\gamma\alpha} = I \delta_{\gamma\alpha} + e^{-i\pi S_{\gamma\alpha}^x}(1-\delta_{\gamma\alpha}),
\end{equation}
and similarly for $U_{\delta\beta}$. Here, $I$ is the identity operator, and the collective operator $S_{\gamma\alpha}^x$ is defined analogously to  $S_{\gamma\alpha}^z$, i.e., $S_{\gamma\alpha}^x = \sum_j \sigma_{\gamma\alpha,x}^j/2$, with $\sigma_{\gamma\alpha,x}^j = \ket{\gamma}_j\bra{\alpha}_j+ \ket{\alpha}_j\bra{\gamma}_j$.
Hence, a sequence of rotate and twist operations provides a fairly universal way to realize BOAT given the ability to perform global single-qudit gates, and entangling operations in a single two-level subspace.

\subsection{Qutrit GHZ states using $^{176}\text{Lu}^+$}

The $^{176}\text{Lu}^+$ ion has previously been identified as a promising clock candidate due to the long-lived $^3D_1$ and $^3D_2$ excited states. In particular, the $^1S_0\leftrightarrow {}^3D_1$ clock transition has an estimated lifetime of $174$ hours~\cite{kaewuam2019spectroscopy,arnold2016observation}, which makes the $^3D_1$ levels highly stable for storing and coherently manipulating quantum information. We consider a chain of $N$ $^{176}\text{Lu}^+$ ions. Denoting the levels as $\ket{{}^3D_1,F,m_F}$, a qutrit can be encoded in three magnetic-field insensitive hyperfine levels $\ket{0}\to \ket{{}^3D_1,6,0}$, $\ket{1}\to \ket{{}^3D_1,7,0}$, $\ket{2}\to \ket{{}^3D_1,8,0}$. The values of $F=6,7,8$ arise because of the nuclear spin $I=7$ for this ion. Single-qudit Raman operations on these qutrits can be implemented by using lasers to address the dipole-allowed $^3D_1\leftrightarrow {}^3P_0$ transitions, one each for the $F=6,7,8$ manifolds. An entangling M\o{}lmer-S\o{}rensen (MS) gate~\cite{molmer1999multiparticle} can be implemented on any pair of levels by utilizing the appropriate pair of Raman lasers to engineer coupling between, say the radial center-of-mass mode, and the relevant electronic levels. The MS gate results in an effective Hamiltonian $\propto S_{\beta\alpha,x}^2$, which can be rotated into the desired form $\propto S_{\beta\alpha,z}^2$ by sandwiching it between global $\pi/2$-pulses along the $y$-axis. By cycling through the three pairs of electronic levels by switching on the appropriate pair of lasers, the BOAT Hamiltonian~(\ref{eqn:ham_qudit}) can be realized.

\subsection{Qutrit and Ququart GHZ states using hybrid $omg$ qudits}

Recent work~\cite{allcock2021omg} has identified the advantages of interconverting between different types of qubit encodings offered by alkaline-earth ions, such as $^{43}\text{Ca}^+$~\cite{benhelm2008experimental} or $^{171}\text{Yb}^+$~\cite{roberts2000observation}, for various quantum operations such as state preparation, gate application and storage. The different qubit encodings enabled by these ions include (i) optical qubits ($o$) consisting of a ground and a metastable level separated by an optical frequency, (ii) metastable qubits ($m$) constituting a pair of long-lived metastable Zeeman or hyperfine levels in an excited state manifold, and (iii) ground-state qubits ($g$) involving a pair of very long-lived Zeeman or hyperfine levels in the ground state manifold. The quantum information processing architecture involving manipulations of, and interconversions between, the different qubit encodings is termed as the $omg$ architecture~\cite{allcock2021omg}. Here, we take inspiration from the ideas underlying this architecture and propose a hybrid encoding of qutrits and ququarts that involves two ground levels and one or two metastable levels, with the ground and metastable manifolds separated by optical frequencies. A combination of lasers addressing the $o$-type transition and microwaves addressing the $m$-type and $g$-type transitions can be used to implement global single-qudit manipulations. Entangling operations on the $g$-type levels can be implemented using MS gates that couple them to auxiliary excited levels~\cite{sackett2000experimental}. In this setting, a serial OAT protocol can be used to realize BOAT by sandwiching OAT operations between swap operations as discussed in Sec.~\ref{sec:serial_oat}. This will lead to the preparation of exotic high-dimensional multipartite GHZ states where the qudits are encoded in energy levels separated by frequencies of different orders of magnitude. 

\section{Entanglement certification}
\label{sec:certification}

Experiments inevitably suffer from non-idealities such as decoherence, and imperfect control and readout. Hence, an important practical question is the quality of the experimental protocol that is required to unambiguously certify that the prepared state indeed features high-dimensional multipartite entanglement. In this section, we will see that the remarkably simple structure of the GHZ state~(\ref{eqn:psi_gen_ghz}) means that measuring a small number of observables--- of order $d^2$ (independent of $N$)--- can reveal plenty of information about the entanglement structure of the prepared state. One way to characterize multipartite entanglement is to study the bipartite entanglement structure of all possible bipartitions $A,\bar{A}$ of the system. For a system of $N$ qudits, a total of $N_b = 2^{N-1}-1$ bipartitions are possible, which we will index with the label $k$~\cite{cadney2014inequalities}. For a given bipartition $k$, the Schmidt decomposition of a pure state can be written as 
\begin{equation}
    \ket{\psi} = \sum_{j=1}^{r_k} \sqrt{c_j} \ket{a_j} \otimes \ket{\bar{a}_j},
\end{equation}
where $r_k$ is the Schmidt rank, $c_j>0$ are the Schmidt coefficients and $\ket{a_j},\ket{\bar{a}_j}$ are a set of mutually orthogonal bases in the $A,\bar{A}$ partitions respectively. Without loss of generality, we assume that $A$ is the partition of smaller size, i.e. $\mathrm{dim}(A)\leq \mathrm{dim}(\bar{A})$. The Schmidt rank $r_k$ is a measure of bipartite entanglement. It is $1$ if $\ket{\psi}$ can be factorized across the $A,\bar{A}$ partitions and is $\mathrm{dim}(A)$ if $\ket{\psi}$ is maximally entangled. 

For a system of $N$ qudits, the multipartite entanglement structure can be characterized by introducing a Schmidt vector $\boldsymbol{r}$, which is an $N_b$-dimensional object specifying the Schmidt ranks of all possible bipartitions of the system. For the $d$-dimensional, $N$-particle GHZ state~(\ref{eqn:psi_gen_ghz}), the Schmidt vector takes an especially simple form,
\begin{equation}
    \boldsymbol{r}_\mathrm{GHZ} = (d,d,d,\ldots),
\end{equation}
i.e., the Schmidt rank is just the qudit dimension $d$ for any bipartition of the $N$-particle system. 

In Refs.~\cite{erhard2018experimental,lierta2022experimental,bao2023verylarge}, the fidelity of the prepared state with the GHZ state was used as a multipartite entanglement witness. In particular, in Refs.~\cite{erhard2018experimental,lierta2022experimental}, it was shown that for a system of $3$ qutrits, measuring a fidelity $\mathcal{F}>2/3$ conclusively proved that the entanglement structure $\boldsymbol{r}_p$ of the prepared state is of the same form as the GHZ state, i.e. $\boldsymbol{r}_p=(3,3,3)$. Extending the analysis beyond the case of $3$ particles and $3$ dimensions is straightforward, and we present this calculation in Appendix~\ref{app:Entanglementstructureandfidelity}. From this analysis, we find the following result: In a system of $N$ qudits with $d$ levels each, measuring a fidelity $\mathcal{F}$ with respect to the $N$-particle $d$-dimensional GHZ state such that 
\begin{equation}
    \mathcal{F} > \frac{d-1}{d}
\end{equation}
is a sufficient condition to certify that the prepared state, typically a density matrix, has at least one pure state in every possible decomposition whose Schmidt vector $\boldsymbol{r}_p$ has \emph{all} elements greater than or equal to $d$. In other words, $\mathcal{F}>(d-1)/d$ rules out entanglement dimensions lower than $d$ in \emph{any} bipartition of the system.

The power of this approach to entanglement certification becomes evident when we consider an example. For a system of, say, $N=10$ qutrits, estimating the fidelity with respect to the corresponding GHZ state requires measuring $9$ observables (see Sec.~\ref{sec:fid_estimation}). An estimate $\mathcal{F}>2/3$ immediately implies that all the $511$ possible bipartitions of the system exhibit at least $3$-dimensional entanglement. Hence, a small number of observables reveal a great deal of information about the entanglement structure of the system.

\section{Fidelity estimation without single-particle control}
\label{sec:fid_estimation}
Although a full reconstruction of the density matrix of a system of $N$ $d$-level qudits requires measurement of $d^N$ observables, the remarkably simple form of the GHZ state means that a measurement of the fidelity requires knowledge of only $d^2$ matrix elements, independent of $N$. These include the populations in the levels $\ket{\alpha}^{\otimes N},\alpha=0,\ldots,d-1$, which can be measured using global manipulations and single-shot readout, and the real and imaginary parts of the coherences between these levels. In previous works with $3$ particles, the latter quantities have been measured by decomposing the observables into expectation values of products of $3$-particle operators, which were then accessed by performing appropriate single-particle rotations on the particles followed by readout~\cite{lierta2022experimental}. However, experiments on some platforms and with a larger number of qudits may lack the ability to address the qudits individually, and instead may be limited to global, identical manipulations of all qudits. In this section, we show that it is nevertheless possible to design a protocol for qutrits to measure the \emph{magnitude} of the coherences between the relevant levels using only global addressing and single-shot readout of only one of the levels. Subsequently, we show how knowledge of the magnitudes can enable placing bounds on the fidelity of the prepared state with the GHZ state.

\subsection{Protocol to measure coherence magnitudes}

\begin{figure*}[!htb]
    \centering
    \includegraphics[width=\textwidth]{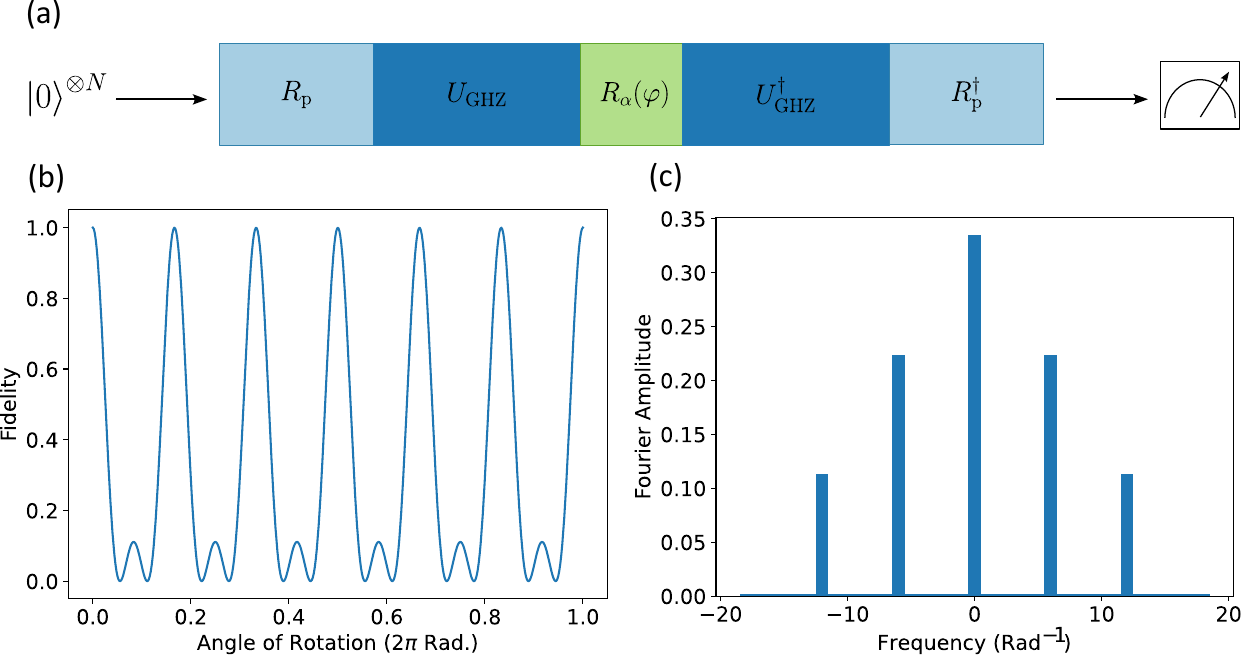}
    \caption{Protocol for estimation of the absolute coherences of the prepared state. (a) Starting in all the atoms in the ground state, we first prepare all the atoms in an equal superposition of the three levels by a global rotation $R_\text{p}$. Then, by applying the BOAT Hamiltonian and proper global rotations, $U_\text{GHZ}$ prepares the GHZ state in the population basis. After that, we perturb the system by applying rotation $R_\alpha(\varphi)$ [Eq.~(\ref{eq:Ralpha})]. Then, reverting the whole process of the state preparation, we measure the fidelity with respect to the initial ground state [Eq.~(\ref{eqn:mqc_fidelity})]. (b) By measuring fidelity at the end of the protocol, we observe oscillations in fidelity with respect to $\varphi$. (c) By performing Fourier analysis, the Fourier component at different frequencies gives information about different coherences of the state [Eq.~\ref{eqn:mqc_coeff}]. In this particular case when $p=2 \text{ and } q =1$ in Eq.~(\ref{eq:Ralpha}), and $N=6$, the non-zero Fourier component at the maximum frequency $2N$ gives the absolute squared value of the coherence between the ground and the second excited level $\abs{\rho_{\{N,0\},\{0,0\}}}^2$.}
    \label{fig:mqc_protocol}
\end{figure*}

Our protocol is a multilevel generalization of the time-reversal scheme demonstrated in Ref.~\cite{gärttner2017measuring}, which was used to measure so-called multiple quantum coherence (MQC) spectra. Without loss of generality, we assume that the state $\ket{0}$ is the `bright' state whose population can be measured, say, via fluorescence. Starting in the state $\ket{0}^{\otimes N}$, the initial state $\ket{\psi(0)}$ [Eq.~(\ref{eqn:qutrit_init})] is prepared using global single-qudit rotations, which we denote by the unitary $R_\text{p}$ in Fig.~\ref{fig:mqc_protocol}(a). Subsequently, the BOAT protocol is applied for time $\chi\tau=2\pi/3$, following which a second global single-qudit rotation is used to align the GHZ state with the population bases, i.e. to rotate the prepared GHZ state to the state~(\ref{eqn:psi_gen_ghz}). We denote the combined application of BOAT and the alignment unitary by $U_\mathrm{GHZ}$. Next, a probe rotation $R_\alpha(\varphi)$ is applied with a variable angle $\varphi$. Finally, the entire state preparation step is `time-reversed' by evolving the system under $U_\mathrm{GHZ}^\dag$ followed by $R_p^\dag$. Physically, the time-reversal step involves changing the sign of the applied Hamiltonian. In trapped ion systems, the sign of the MS gate or the LS gate can be reversed by changing the detuning of the spin-dependent force from the center-of-mass mode~\cite{garttner2018relating}. The sign of single-particle operations can also be reversed by suitable phase shifts of the applied lasers or microwave fields. Finally, the population in the state $\ket{0}$ is measured for all the ions and the fidelity $\mathcal{F}(\varphi)$ with the state $\ket{0}^{\otimes N}$ is estimated as the fraction of experimental shots where all the ions are measured to be in $\ket{0}$. The requirement for individual-ion readout may also be relaxed: The number of ions in the bright state has previously been measured in the qubit case by collecting the total fluorescence from all the ions and comparing it with an independent calibration of the photons collected per ion~\cite{gärttner2017measuring}. If the variable phase in $R_\alpha(\varphi)$ is taken to be $\varphi=0$, then the final state is exactly the same as the initial state and $\mathcal{F}(0)=1$. On the other hand, a plot of $\mathcal{F}(\varphi)$ verus $\varphi$ displays oscillations, which contain information about coherences between the different qudit levels. 

To see this, we denote the initial state as $\rho_0 = \ket{0}^{\otimes N}\bra{0}^{\otimes N}$, using which the final state $\rho_f$ prior to measurement can be expressed as [see Fig.~\ref{fig:mqc_protocol}(a)]
\begin{equation}
    \rho_f = R_p^\dag U_\mathrm{GHZ}^\dag R_\alpha(\varphi) U_\mathrm{GHZ} R_p\rho_0 R_p^\dag U_\mathrm{GHZ}^\dag R_\alpha^\dag (\varphi) U_\mathrm{GHZ} R_p.
\end{equation}
The fidelity with the initial state pure state $\rho_0$ is then given by 
\begin{align}
    \mathcal{F}(\varphi) &= \mathrm{Tr} [\rho_f \rho_0] 
    = \mathrm{Tr}[R_\alpha(\varphi) \rho R_\alpha^\dag \rho],
    \label{eqn:mqc_fidelity}
\end{align}
where $\rho$ is the state of the system before the application of the variable rotation $R_\alpha(\varphi)$, given by 
\begin{equation}
    \rho = U_\mathrm{GHZ} R_p\rho_0 R_p^\dag U_\mathrm{GHZ}^\dag.
\end{equation}
Equation~(\ref{eqn:mqc_fidelity})  thus implies that $\mathcal{F}(\varphi)$ is essentially a measure of the overlap between the state prepared by the BOAT protocol and a rotated version of itself.

We consider the rotation operator to be of the form 
\begin{equation}
    R_\alpha(\varphi) = e^{-i\varphi(p S_{11} + q S_{22})},
    \label{eq:Ralpha}
\end{equation}
where $S_{\alpha\alpha} = \sum_j \ket{\alpha}_j\bra{\alpha}_j$ captures the number of atoms in $\ket{\alpha}$ and $p,q$ are integers. Physically, such a rotation can be implemented by inducing ac Stark shifts via off-resonant lasers that couple the levels $\ket{1}$ and $\ket{2}$ to auxiliary levels. Expressing $\rho$ in the $\ket{N,\boldsymbol{l}}$ basis, rotation by $R_\alpha(\varphi)$ leads to 
\begin{equation}
    R_\alpha(\varphi) \rho R_\alpha^\dag = \sum_{\boldsymbol{l},\boldsymbol{l}'}\rho_{\boldsymbol{l},\boldsymbol{l}'} e^{-i\varphi(p(l_1-l_1') + q(l_2-l_2'))} \ket{\boldsymbol{l}}\bra{\boldsymbol{l}'}.
    \label{eqn:r_phi}
\end{equation}
Introducing the index $m$ as 
\begin{equation}
    m = p(l_1-l_1') + q(l_2-l_2'),
    \label{eqn:m_ind}
\end{equation}
$\mathcal{F}(\varphi)$ can be expressed as 
\begin{equation}
    \mathcal{F}(\varphi) = \sum_{m} \mathcal{I}_m e^{-im\varphi},
\end{equation}
where the coefficients $\mathcal{I}_m$ form the so-called multiple quantum coherence spectrum~\cite{gärttner2017measuring,garttner2018relating,lewisswan2019unifying} and can be extracted from $\mathcal{F}(\varphi)$ through a Fourier transform:
\begin{equation}
    \mathcal{I}_m = \frac{1}{2\pi}\int_0^{2\pi}d\varphi \mathcal{F}(\varphi) =  \sum_{\substack{\boldsymbol{l},\boldsymbol{l}'\\ p(l_1-l_1')+q(l_2-l_2')=m}} \abs{\rho_{\boldsymbol{l},\boldsymbol{l}'}}^2.
    \label{eqn:mqc_coeff}
\end{equation}
Hence, the MQC coefficients contain information about the sum of squared magnitudes of all matrix elements between bases $\boldsymbol{l},\boldsymbol{l}'$ which are related according to Eq.~(\ref{eqn:m_ind}).
In particular, we are interested in coherences between the GHZ bases states $\ket{\alpha}^{\otimes N},\alpha=0,1,2$. These coherences can be isolated in the above protocol using specific choices of the integers $p,q$ in Eq.~(\ref{eqn:r_phi}) and measuring the coefficient $\mathcal{I}_{2N}$:
\begin{enumerate}
    \item $p=2,\;q=1$: In this case, $\mathcal{I}_{2N} = \abs{\rho_{\{N,0\},\{0,0\}}}^2$, which is the magnitude of the coherence between $\ket{0}^{\otimes N}$ and  $\ket{1}^{\otimes N}$.
    \item $p=1,\;q=2$: In this case, $\mathcal{I}_{2N} = \abs{\rho_{\{0,N\},\{0,0\}}}^2$, which is the magnitude of the coherence between $\ket{0}^{\otimes N}$ and  $\ket{2}^{\otimes N}$.
    \item $p=-1,\;q=1$: In this case, $\mathcal{I}_{2N} = \abs{\rho_{\{0,N\},\{N,0\}}}^2$, which is the magnitude of the coherence between $\ket{1}^{\otimes N}$ and  $\ket{2}^{\otimes N}$.    
\end{enumerate}
Therefore, repeating the above protocol with three difference choices of the rotation operator $R_\alpha(\varphi)$ enables measurement of the coherence magnitudes between the GHZ bases without the requirement for individual qudit addressing. 

As an example, Figure~\ref{fig:mqc_protocol}(b) shows a plot of $\mathcal{F}(\varphi)$ versus $\varphi$ for the case of $p=2,q=1$ and for a system of $N=6$ qutrits. Taking a Fourier transform of this curve according to Eq.~(\ref{eqn:mqc_coeff}) gives the MQC spectrum, $\mathcal{I}_m$ versus $m$, shown in Fig.~\ref{fig:mqc_protocol}(c). In particular, the coefficient $\mathcal{I}_{\pm 12}$ corresponding to $m=\pm 2N$ has a value $\mathcal{I}_{\pm 12}=1/9\approx 0.11$, from which we can immediately infer the magnitude $\abs{\rho_{\{N,0\},\{0,0\}}}=\sqrt{\mathcal{I}_{\pm 12}} = 1/3$, which is the expected value for a $3$-dimensional GHZ state.\\

We now briefly comment on the impact of decoherence on the above protocol. The central question is whether Eq.~(\ref{eqn:mqc_fidelity}) is still valid in the presence of decoherence, i.e., whether measuring the fidelity of the initial and final states is equivalent to measuring the fidelity of the prepared state and its rotated version. The authors of Ref.~\cite{garttner2018relating} showed that the equivalence holds in the presence of dephasing processes, whereas for state-changing processes, their analysis essentially implies that the equivalence holds as long as the forward and backward rates between pairs of levels are approximately the same, which ensures a time-reversal symmetry between the two processes. 

\subsection{Bounds on fidelity}

Given an estimate of only the coherence magnitudes, rather than the real and imaginary parts of the coherences, it is still possible to place bounds on the fidelity $\mathcal{F}$ of the prepared state $\rho$ with the family of GHZ states such that 
\begin{equation}
    \mathcal{F}_\mathrm{l} \leq \mathcal{F} \leq \mathcal{F}_\mathrm{u}.    
\end{equation}
To derive the upper and lower bounds $\mathcal{F}_\mathrm{u}$ and $\mathcal{F}_\mathrm{l}$ , we first note that a general $3$-dimensional GHZ state has two free phases $\phi_1,\phi_2$, since it is of the form 
\begin{equation}
\ket{\psi}_{\text{t}} = \frac{1}{\sqrt{3}} (\ket{0}^{\otimes N} + e^{i\phi_1}\ket{1}^{\otimes N} + e^{i\phi_2}\ket{2}^{\otimes N}).    
\end{equation}
Here, the subscript `t' denotes that this is the target state. On the other hand, given measurement of the populations and coherence magnitudes, our knowledge of the prepared state in the $3\times 3$ subspace spanned by $\ket{\alpha}^{\otimes N}$ is given by a density matrix parametrized by three unknown phases:
\begin{align}
\rho = \begin{pmatrix}
\rho_{00} &  |\rho_{01}| e^{i\theta_1} & |\rho_{02}| e^{i\theta_2} \\ 
|\rho_{01}| e^{-i\theta_1} &  \rho_{11} & |\rho_{12} | e^{i\theta_3} \\ 
|\rho_{02}| e^{-i\theta_2} &  |\rho_{12}| e^{-i\theta_3} & \rho_{22} \\ 
\end{pmatrix}.
\label{eqn:dmat}
\end{align}
Hence, the fidelity between the prepared state and the target state can be expressed as 
\begin{align}
\begin{split}
\mathcal{F} &= \frac{\rho_{00} + \rho_{11} + \rho_{22}}{3} + 
\frac{2}{3} \left( |\rho_{01}| \cos({\theta_1-\phi_1}) \right. \\&\left.+ |\rho_{02}| \cos({\theta_2-\phi_2}) + |\rho_{12}| \cos({\theta_3-(\phi_1 - \phi_2)})   \right).
\end{split}
\end{align}
Without loss of generality, let us assume that $\abs{\rho_{01}}\geq \abs{\rho_{02}}\geq \abs{\rho_{12}}$~\footnote{If this is not satisfied, we can always relabel the levels to meet this condition.}. Since $\phi_1,\phi_2$ are free parameters of the target GHZ state, they can always be chosen to satisfy $\phi_1=\theta_1$ and $\phi_2=\theta_2$. Hence, the fidelity takes the form 
\begin{equation}
    \mathcal{F} = \frac{\rho_{00} + \rho_{11} + \rho_{22}}{3} + 
\frac{2}{3} \left( |\rho_{01}| + |\rho_{02}|  + |\rho_{12}|\cos\theta   \right),
\label{eqn:fid_costheta}
\end{equation}
where $\theta=\theta_3-(\phi_1-\phi_2)$ is an unknown phase.

An upperbound on $\mathcal{F}$ is easily obtained by setting $\theta=0$:
\begin{align}
\begin{split}
\mathcal{F}_\mathrm{u} &= \frac{\rho_{00} + \rho_{11} + \rho_{22}}{3} + 
\frac{2}{3} \left( |\rho_{01}| + |\rho_{02}|  + |\rho_{12}|   \right).
\end{split}
\label{eqn:fid_ub}
\end{align}
When the prepared state is a perfect GHZ state, all the $6$ quantities in Eq.~(\ref{eqn:fid_ub}) equal $1/3$, and hence we recover the expected result that $\mathcal{F}_\mathrm{u}=1$.

To obtain the lower bound, the central observation is that $\theta$ cannot be arbitrary; it is constrained by the physicality requirement that the density matrix~(\ref{eqn:dmat}) must be positive semi-definite. Given the populations and coherence magnitudes, the condition of $\rho\geq 0$ leads to the constraint $\cos\theta\geq s$, where 
\begin{equation}
     s = \frac{\left|\rho_{12}\right|^2 \rho_{00}+\left|\rho_{01}\right|^2 \rho_{22}+\left|\rho_{02}\right|^2 \rho_{11} - \rho_{00} \rho_{11} \rho_{22}}{2 \left|\rho_{01} \rho_{12} \rho_{20}\right|}.
     \label{eqn:s_fid_lb}
\end{equation}
The derivation of this expression is provided in Appendix~\ref{app:fid_lb}. Hence, a lower bound on $\mathcal{F}$ is given by 
\begin{equation}
        \mathcal{F}_\mathrm{l} = \frac{\rho_{00} + \rho_{11} + \rho_{22}}{3} + 
\frac{2}{3} \left( |\rho_{01}| + |\rho_{02}|  + |\rho_{12}|s   \right).
\label{eqn:fid_lb}
\end{equation}
In the ideal scenario where the prepared state is identical to a GHZ state, we find that $s=1$, and the two bounds coincide, i.e. $\mathcal{F}_\mathrm{l}=\mathcal{F}_\mathrm{u}=1$.

The above analysis implies that an experiment must target to achieve a lower bound $\mathcal{F}_\mathrm{l}\geq 2/3$ in order to demonstrate high-dimensional multipartite entanglement in the sense described in Sec.~\ref{sec:certification}.

\section{Conclusion and outlook}
\label{sec:conclusion}

\appendix

We have proposed a scalable path to prepare and certify high-dimensional multipartite entangled states outside of photonic systems. Our scheme, which we call as balanced one-axis twisting (BOAT), is a multilevel generalization of the OAT interaction and enables the preparation of generalized GHZ states in $3$ and $4$ dimensions and for any number of particles. We also discussed how measuring the fidelity of the prepared state enables the certification of high-dimensional multipartite entanglement. Furthermore, we showed how the fidelity can be bounded using a time-reversal protocol that relies only on global control, followed by single-shot readout of one of the levels.  

Although our experimental proposals are targeted at trapped ion systems, the BOAT protocol can also be realized on other platforms where all-to-all coupling is available or can be engineered. For instance, the Ising type interactions assumed between pairs of levels in this work can also be replaced with a spin-exchange type interaction, which is natively available in atom-cavity systems~\cite{hu2017vacuum,norcia2018cavity} and which has also been demonstrated in circuit QED systems~\cite{xu2022metrological,hazra2021ringresonator}. Alternatively, all-to-all Ising type interactions can also be realized with finite range interactions, e.g., in ensembles of Rydberg atoms that are placed within a single Rydberg blockade radius~\cite{cao2024multiqubitgates}.

The BOAT scheme relies on repeated application of effective qubit-type entangling gates coupling pairs of levels embedded in a qudit system. Future work can explore the preparation of high-dimensional GHZ states  using genuine multilevel entangling gates, which have been demonstrated in trapped ion systems~\cite{hrmo2023native}. Variational approaches~\cite{kaubruegger2021quantum,marciniak2022optimal} to state preparation can also be explored, which could help in minimizing the state preparation time, and could also reveal paths to prepare GHZ states in $d>4$ dimensions. Finally, in this work, we focused specifically on the preparation of GHZ states. More generally, it will be interesting to explore the dynamical generation of multilevel, multi-headed cat states and high-dimensional spin squeezing~\cite{sundar2024driven} under the BOAT protocol, which may be of interest, e.g., for multiparameter quantum metrology.

\begin{acknowledgments}
We thank Ana Maria Rey, John Bollinger and Vaibhav Madhok for helpful discussions. H.V.U. acknowledges the support of
Ministry of Education, Government of India.
T.R.T. acknowledges funding from the 
Australian Research Council (FT220100359) and the Sydney Horizon Fellowship. B.S. acknowledges support from Ministry of Electronics and Information Technology, Govt of India under the Centre for Excellence in Quantum Technology grant. A.S. acknowledges the support of a C.V. Raman Postdoctoral Fellowship from the Indian Institute of Science. 
\end{acknowledgments}

\section{Entanglement structure and fidelity}\label{app:Entanglementstructureandfidelity}

In this appendix, we will show that if the observed fidelity with the target GHZ state~(\ref{eqn:psi_gen_ghz}) is greater than $(d-1)/d$, then our final state, generally being a mixed state, must necessarily have at least one pure state whose Schmidt vector has all elements greater than or equal to $d$. The Schmidt vector of a GHZ state has all entries equal to $d$. Now, we  consider a state $\ket{\phi}$ whose Schmidt vector has one entry, indexed by $k$, equal to $d-1$, and all the rest equal to $d$.  To find the maximum fidelity $\mathcal{F}_\mathrm{th}$ that can be achieved by a state with this Schmidt vector with respect to a GHZ state $\ket{\Psi}$, we maximize over all the possible states $\ket{\phi}$, i.e.,
\begin{align}
\mathcal{F}_\mathrm{th} = \text{max}_{\ket{\phi}}\left(\text{Tr}\left(\ket{\phi}\bra{\phi}\ket{\Psi}\bra{\Psi}\right)\right).   
\end{align} \\
As the fidelity is convex with respect to the states, we can restrict the analysis to pure states.

The central idea of this calculation is to consider the bipartition with the lowest Schmidt rank. We express $\ket{\phi}$ across the  $k^\text{th}$ bipartition using its Schmidt decomposition as $\ket{\phi}=\sum_{i=0}^{d-2}c_i \ket{u_i}\ket{v_i}$. For the GHZ state, the corresponding decomposition is $\ket{\Psi}=\sum_{i=0}^{d-1} \frac{1}{\sqrt{d}} \ket{i}\ket{i}$, where a state $\ket{i}$ for a partition means that all the atoms in that partition are in the $i^\text{th}$ level. Now, for $\ket{\phi}$, we can write a $d-2$ rank projection operator $P = \sum_{i=0}^{d-2}\ket{\lambda_i}\bra{\lambda_i}\otimes I$ where $\braket{\lambda_i | \lambda_j} = \delta_{ij}$ for all $i,j \in [0, 1, \ldots, d-2]$, such that $P^2 = P$ and 

$$
P\ket{\phi}\bra{\phi}=\ket{\phi}\bra{\phi}P=\ket{\phi}\bra{\phi}.
$$
Note that this operator only acts on the first partition. We then have, 
\begin{align}
\mathcal{F}_\mathrm{th} & = \text{max}_{\ket{\phi}}\bigg(\text{Tr}\left(P\ket{\phi}\bra{\phi}P\ket{\Psi}\bra{\Psi}\right)\bigg)\nonumber\\    
& = \text{max}_{\ket{\phi}}\bigg(\text{Tr}\left(\ket{\phi}\bra{\phi}P\ket{\Psi}\bra{\Psi}P\right)\bigg).
\end{align}
Using the triangle inequality, we have

\begin{align}
\mathcal{F}_\mathrm{th} \le \text{max}_{\ket{\phi}}\bigg(\text{Tr}\left(\ket{\phi}\bra{\phi}\right)\text{Tr}\left(P\ket{\Psi}\bra{\Psi}P\right)\bigg).
\end{align}
The first term will be 1, and maximizing over $\ket{\phi}$ now translates to maximizing over $P$:
\begin{align}
\mathcal{F}_\mathrm{th} \le \text{max}_P\text{Tr}\left(\ket{\Psi}\bra{\Psi}P\right). 
\end{align}
Writing $\ket{\Psi}$ as $\sum_{i=0}^{d-1} \frac{1}{\sqrt{d}} \ket{i}\ket{i}$ and noting that all the $\ket{i}$ are orthonormal, we have
\begin{align}
\mathcal{F}_\mathrm{th} &\le \frac{1}{d}\text{max}_{P} \text{Tr} \left(\sum_{i=0}^{d-1}\bra{i} P \ket{i}\right).
\end{align}
This value will be maximum when $P$ acts in the subspace generated by states $\ket{i}$, $i\in \{0, 1, \ldots, d-1\}$, in which case 
\begin{align}
\mathcal{F}_\mathrm{th} \le \frac{1}{d}\text{max}_{P}\text{Tr}\left(   P\right).
\end{align}
As $P$ is a $d-1$ dimensional projection operator $\text{Tr}(P) = d-1$, which gives
\begin{align}
\mathcal{F}_\mathrm{th} \le \frac{d-1}{d}.
\end{align}

Hence, the maximum fidelity that can be achieved with respect to a GHZ state by a state of lower entanglement structure is $(d-1)/d$. Therefore, if, in our experiments, we find a fidelity greater than this value, then our final mixed state must have at least one pure state whose Schmidt vector has all elements $\geq d$.

\section{Derivation of fidelity lower bound}
\label{app:fid_lb}

In this appendix, we derive the lower bound on the fidelity of the prepared state and the family of GHZ states, Eq.~(\ref{eqn:fid_lb}). The value of $\cos\theta$ in Eq.~(\ref{eqn:fid_costheta}) cannot be chosen arbitrarily, as it is constrained by the requirement that all the eigenvalues of the density matrix $\rho$~(\ref{eqn:dmat}) must be nonnegative.  The eigenvalue equation for $\rho$ in variable $\lambda$ is 
\begin{align}
\begin{split}& \lambda^3-\lambda^2\left(\rho_{00}+\rho_{11}+\rho_{22}\right)  +\lambda\Big(\rho_{00} \rho_{11}+\rho_{11} \rho_{22}+\rho_{00} \rho_{22}  \\& -\left.\left(\left|\rho_{01}\right|^2+\left|\rho_{12}\right|^2+\left|\rho_{02}\right|^2\right)\right)  - \rho_{00} \rho_{11} \rho_{22} +\left|\rho_{22}\right|^2 \rho_{00} \\&+\left|\rho_{01}\right|^2 \rho_{22}+\left|\rho_{02}\right|^2 \rho_{11} - \rho_{01} \rho_{12} \rho_{20}  -\rho_{02} \rho_{10} \rho_{21}=0. 
\end{split}
\end{align}
For any equation of the form $y(x) = x^3 + bx^2 + cx + d$ to have three nonnegative roots, a necessary condition is that $d\le0$, which leads to the condition
\begin{align}
\begin{split}
- \rho_{00} \rho_{11} \rho_{22}+\left|\rho_{12}\right|^2 \rho_{00}+\left|\rho_{01}\right|^2 \rho_{22}+\left|\rho_{02}\right|^2 \rho_{11} \\- \rho_{01} \rho_{12} \rho_{20}  -\rho_{02} \rho_{10} \rho_{21} \le 0. 
\end{split}
\end{align}
Writing $\rho_{02} \rho_{10} \rho_{21} = |\rho_{01} \rho_{12} \rho_{20}| e^{i\theta}$ and rearranging terms leads to  the condition $\cos\theta\geq s$, where $s$ is given by Eq.~(\ref{eqn:s_fid_lb}).


\begin{thebibliography}{48}%
\makeatletter
\providecommand \@ifxundefined [1]{%
 \@ifx{#1\undefined}
}%
\providecommand \@ifnum [1]{%
 \ifnum #1\expandafter \@firstoftwo
 \else \expandafter \@secondoftwo
 \fi
}%
\providecommand \@ifx [1]{%
 \ifx #1\expandafter \@firstoftwo
 \else \expandafter \@secondoftwo
 \fi
}%
\providecommand \natexlab [1]{#1}%
\providecommand \enquote  [1]{``#1''}%
\providecommand \bibnamefont  [1]{#1}%
\providecommand \bibfnamefont [1]{#1}%
\providecommand \citenamefont [1]{#1}%
\providecommand \href@noop [0]{\@secondoftwo}%
\providecommand \href [0]{\begingroup \@sanitize@url \@href}%
\providecommand \@href[1]{\@@startlink{#1}\@@href}%
\providecommand \@@href[1]{\endgroup#1\@@endlink}%
\providecommand \@sanitize@url [0]{\catcode `\\12\catcode `\$12\catcode `\&12\catcode `\#12\catcode `\^12\catcode `\_12\catcode `\%12\relax}%
\providecommand \@@startlink[1]{}%
\providecommand \@@endlink[0]{}%
\providecommand \url  [0]{\begingroup\@sanitize@url \@url }%
\providecommand \@url [1]{\endgroup\@href {#1}{\urlprefix }}%
\providecommand \urlprefix  [0]{URL }%
\providecommand \Eprint [0]{\href }%
\providecommand \doibase [0]{https://doi.org/}%
\providecommand \selectlanguage [0]{\@gobble}%
\providecommand \bibinfo  [0]{\@secondoftwo}%
\providecommand \bibfield  [0]{\@secondoftwo}%
\providecommand \translation [1]{[#1]}%
\providecommand \BibitemOpen [0]{}%
\providecommand \bibitemStop [0]{}%
\providecommand \bibitemNoStop [0]{.\EOS\space}%
\providecommand \EOS [0]{\spacefactor3000\relax}%
\providecommand \BibitemShut  [1]{\csname bibitem#1\endcsname}%
\let\auto@bib@innerbib\@empty
\bibitem [{\citenamefont {Bao}\ \emph {et~al.}(2023)\citenamefont {Bao}, \citenamefont {Fu}, \citenamefont {Pramanik}, \citenamefont {Mao}, \citenamefont {Chi}, \citenamefont {Cao}, \citenamefont {Zhai}, \citenamefont {Mao}, \citenamefont {Dai}, \citenamefont {Chen}, \citenamefont {Jia}, \citenamefont {Zhao}, \citenamefont {Zheng}, \citenamefont {Tang}, \citenamefont {Li}, \citenamefont {Luo}, \citenamefont {Wang}, \citenamefont {Yang}, \citenamefont {Peng}, \citenamefont {Liu}, \citenamefont {Dai}, \citenamefont {He}, \citenamefont {Muthali}, \citenamefont {Oxenl{\o}we}, \citenamefont {Vigliar}, \citenamefont {Paesani}, \citenamefont {Hou}, \citenamefont {Santagati}, \citenamefont {Silverstone}, \citenamefont {Laing}, \citenamefont {Thompson}, \citenamefont {O'Brien}, \citenamefont {Ding}, \citenamefont {Gong},\ and\ \citenamefont {Wang}}]{bao2023verylarge}%
  \BibitemOpen
  \bibfield  {author} {\bibinfo {author} {\bibfnamefont {J.}~\bibnamefont {Bao}}, \bibinfo {author} {\bibfnamefont {Z.}~\bibnamefont {Fu}}, \bibinfo {author} {\bibfnamefont {T.}~\bibnamefont {Pramanik}}, \bibinfo {author} {\bibfnamefont {J.}~\bibnamefont {Mao}}, \bibinfo {author} {\bibfnamefont {Y.}~\bibnamefont {Chi}}, \bibinfo {author} {\bibfnamefont {Y.}~\bibnamefont {Cao}}, \bibinfo {author} {\bibfnamefont {C.}~\bibnamefont {Zhai}}, \bibinfo {author} {\bibfnamefont {Y.}~\bibnamefont {Mao}}, \bibinfo {author} {\bibfnamefont {T.}~\bibnamefont {Dai}}, \bibinfo {author} {\bibfnamefont {X.}~\bibnamefont {Chen}}, \bibinfo {author} {\bibfnamefont {X.}~\bibnamefont {Jia}}, \bibinfo {author} {\bibfnamefont {L.}~\bibnamefont {Zhao}}, \bibinfo {author} {\bibfnamefont {Y.}~\bibnamefont {Zheng}}, \bibinfo {author} {\bibfnamefont {B.}~\bibnamefont {Tang}}, \bibinfo {author} {\bibfnamefont {Z.}~\bibnamefont {Li}}, \bibinfo {author} {\bibfnamefont {J.}~\bibnamefont {Luo}}, \bibinfo {author} {\bibfnamefont
  {W.}~\bibnamefont {Wang}}, \bibinfo {author} {\bibfnamefont {Y.}~\bibnamefont {Yang}}, \bibinfo {author} {\bibfnamefont {Y.}~\bibnamefont {Peng}}, \bibinfo {author} {\bibfnamefont {D.}~\bibnamefont {Liu}}, \bibinfo {author} {\bibfnamefont {D.}~\bibnamefont {Dai}}, \bibinfo {author} {\bibfnamefont {Q.}~\bibnamefont {He}}, \bibinfo {author} {\bibfnamefont {A.~L.}\ \bibnamefont {Muthali}}, \bibinfo {author} {\bibfnamefont {L.~K.}\ \bibnamefont {Oxenl{\o}we}}, \bibinfo {author} {\bibfnamefont {C.}~\bibnamefont {Vigliar}}, \bibinfo {author} {\bibfnamefont {S.}~\bibnamefont {Paesani}}, \bibinfo {author} {\bibfnamefont {H.}~\bibnamefont {Hou}}, \bibinfo {author} {\bibfnamefont {R.}~\bibnamefont {Santagati}}, \bibinfo {author} {\bibfnamefont {J.~W.}\ \bibnamefont {Silverstone}}, \bibinfo {author} {\bibfnamefont {A.}~\bibnamefont {Laing}}, \bibinfo {author} {\bibfnamefont {M.~G.}\ \bibnamefont {Thompson}}, \bibinfo {author} {\bibfnamefont {J.~L.}\ \bibnamefont {O'Brien}}, \bibinfo {author} {\bibfnamefont
  {Y.}~\bibnamefont {Ding}}, \bibinfo {author} {\bibfnamefont {Q.}~\bibnamefont {Gong}},\ and\ \bibinfo {author} {\bibfnamefont {J.}~\bibnamefont {Wang}},\ }\bibfield  {title} {\bibinfo {title} {Very-large-scale integrated quantum graph photonics},\ }\href {https://doi.org/10.1038/s41566-023-01187-z} {\bibfield  {journal} {\bibinfo  {journal} {Nature Photonics}\ }\textbf {\bibinfo {volume} {17}},\ \bibinfo {pages} {573} (\bibinfo {year} {2023})}\BibitemShut {NoStop}%
\bibitem [{\citenamefont {Xing}\ \emph {et~al.}(2023)\citenamefont {Xing}, \citenamefont {Hu}, \citenamefont {Guo}, \citenamefont {Liu}, \citenamefont {Li},\ and\ \citenamefont {Guo}}]{xing2023preparation}%
  \BibitemOpen
  \bibfield  {author} {\bibinfo {author} {\bibfnamefont {W.-B.}\ \bibnamefont {Xing}}, \bibinfo {author} {\bibfnamefont {X.-M.}\ \bibnamefont {Hu}}, \bibinfo {author} {\bibfnamefont {Y.}~\bibnamefont {Guo}}, \bibinfo {author} {\bibfnamefont {B.-H.}\ \bibnamefont {Liu}}, \bibinfo {author} {\bibfnamefont {C.-F.}\ \bibnamefont {Li}},\ and\ \bibinfo {author} {\bibfnamefont {G.-C.}\ \bibnamefont {Guo}},\ }\bibfield  {title} {\bibinfo {title} {Preparation of multiphoton high-dimensional ghz states},\ }\href {https://doi.org/10.1364/OE.494850} {\bibfield  {journal} {\bibinfo  {journal} {Opt. Express}\ }\textbf {\bibinfo {volume} {31}},\ \bibinfo {pages} {24887} (\bibinfo {year} {2023})}\BibitemShut {NoStop}%
\bibitem [{\citenamefont {Humphreys}\ \emph {et~al.}(2013)\citenamefont {Humphreys}, \citenamefont {Barbieri}, \citenamefont {Datta},\ and\ \citenamefont {Walmsley}}]{humphreys2013quantum}%
  \BibitemOpen
  \bibfield  {author} {\bibinfo {author} {\bibfnamefont {P.~C.}\ \bibnamefont {Humphreys}}, \bibinfo {author} {\bibfnamefont {M.}~\bibnamefont {Barbieri}}, \bibinfo {author} {\bibfnamefont {A.}~\bibnamefont {Datta}},\ and\ \bibinfo {author} {\bibfnamefont {I.~A.}\ \bibnamefont {Walmsley}},\ }\bibfield  {title} {\bibinfo {title} {Quantum enhanced multiple phase estimation},\ }\href {https://doi.org/10.1103/PhysRevLett.111.070403} {\bibfield  {journal} {\bibinfo  {journal} {Phys. Rev. Lett.}\ }\textbf {\bibinfo {volume} {111}},\ \bibinfo {pages} {070403} (\bibinfo {year} {2013})}\BibitemShut {NoStop}%
\bibitem [{\citenamefont {Trényi}\ \emph {et~al.}(2024)\citenamefont {Trényi}, \citenamefont {Árpád Lukács}, \citenamefont {Horodecki}, \citenamefont {Horodecki}, \citenamefont {Vértesi},\ and\ \citenamefont {Tóth}}]{trenyi2024activation}%
  \BibitemOpen
  \bibfield  {author} {\bibinfo {author} {\bibfnamefont {R.}~\bibnamefont {Trényi}}, \bibinfo {author} {\bibnamefont {Árpád Lukács}}, \bibinfo {author} {\bibfnamefont {P.}~\bibnamefont {Horodecki}}, \bibinfo {author} {\bibfnamefont {R.}~\bibnamefont {Horodecki}}, \bibinfo {author} {\bibfnamefont {T.}~\bibnamefont {Vértesi}},\ and\ \bibinfo {author} {\bibfnamefont {G.}~\bibnamefont {Tóth}},\ }\bibfield  {title} {\bibinfo {title} {Activation of metrologically useful genuine multipartite entanglement},\ }\href {https://doi.org/10.1088/1367-2630/ad1e93} {\bibfield  {journal} {\bibinfo  {journal} {New Journal of Physics}\ }\textbf {\bibinfo {volume} {26}},\ \bibinfo {pages} {023034} (\bibinfo {year} {2024})}\BibitemShut {NoStop}%
\bibitem [{\citenamefont {Chi}\ \emph {et~al.}(2022)\citenamefont {Chi}, \citenamefont {Huang}, \citenamefont {Zhang}, \citenamefont {Mao}, \citenamefont {Zhou}, \citenamefont {Chen}, \citenamefont {Zhai}, \citenamefont {Bao}, \citenamefont {Dai}, \citenamefont {Yuan}, \citenamefont {Zhang}, \citenamefont {Dai}, \citenamefont {Tang}, \citenamefont {Yang}, \citenamefont {Li}, \citenamefont {Ding}, \citenamefont {Oxenløwe}, \citenamefont {Thompson}, \citenamefont {O’Brien}, \citenamefont {Li}, \citenamefont {Gong},\ and\ \citenamefont {Wang}}]{chi_programmable_2022}%
  \BibitemOpen
  \bibfield  {author} {\bibinfo {author} {\bibfnamefont {Y.}~\bibnamefont {Chi}}, \bibinfo {author} {\bibfnamefont {J.}~\bibnamefont {Huang}}, \bibinfo {author} {\bibfnamefont {Z.}~\bibnamefont {Zhang}}, \bibinfo {author} {\bibfnamefont {J.}~\bibnamefont {Mao}}, \bibinfo {author} {\bibfnamefont {Z.}~\bibnamefont {Zhou}}, \bibinfo {author} {\bibfnamefont {X.}~\bibnamefont {Chen}}, \bibinfo {author} {\bibfnamefont {C.}~\bibnamefont {Zhai}}, \bibinfo {author} {\bibfnamefont {J.}~\bibnamefont {Bao}}, \bibinfo {author} {\bibfnamefont {T.}~\bibnamefont {Dai}}, \bibinfo {author} {\bibfnamefont {H.}~\bibnamefont {Yuan}}, \bibinfo {author} {\bibfnamefont {M.}~\bibnamefont {Zhang}}, \bibinfo {author} {\bibfnamefont {D.}~\bibnamefont {Dai}}, \bibinfo {author} {\bibfnamefont {B.}~\bibnamefont {Tang}}, \bibinfo {author} {\bibfnamefont {Y.}~\bibnamefont {Yang}}, \bibinfo {author} {\bibfnamefont {Z.}~\bibnamefont {Li}}, \bibinfo {author} {\bibfnamefont {Y.}~\bibnamefont {Ding}}, \bibinfo {author} {\bibfnamefont {L.~K.}\
  \bibnamefont {Oxenløwe}}, \bibinfo {author} {\bibfnamefont {M.~G.}\ \bibnamefont {Thompson}}, \bibinfo {author} {\bibfnamefont {J.~L.}\ \bibnamefont {O’Brien}}, \bibinfo {author} {\bibfnamefont {Y.}~\bibnamefont {Li}}, \bibinfo {author} {\bibfnamefont {Q.}~\bibnamefont {Gong}},\ and\ \bibinfo {author} {\bibfnamefont {J.}~\bibnamefont {Wang}},\ }\bibfield  {title} {\bibinfo {title} {A programmable qudit-based quantum processor},\ }\href {https://doi.org/10.1038/s41467-022-28767-x} {\bibfield  {journal} {\bibinfo  {journal} {Nature Communications}\ }\textbf {\bibinfo {volume} {13}},\ \bibinfo {pages} {1166} (\bibinfo {year} {2022})},\ \bibinfo {note} {publisher: Nature Publishing Group}\BibitemShut {NoStop}%
\bibitem [{\citenamefont {Ryu}\ \emph {et~al.}(2013)\citenamefont {Ryu}, \citenamefont {Lee}, \citenamefont {\ifmmode~\dot{Z}\else \.{Z}\fi{}ukowski},\ and\ \citenamefont {Lee}}]{ryu2013greenberger}%
  \BibitemOpen
  \bibfield  {author} {\bibinfo {author} {\bibfnamefont {J.}~\bibnamefont {Ryu}}, \bibinfo {author} {\bibfnamefont {C.}~\bibnamefont {Lee}}, \bibinfo {author} {\bibfnamefont {M.}~\bibnamefont {\ifmmode~\dot{Z}\else \.{Z}\fi{}ukowski}},\ and\ \bibinfo {author} {\bibfnamefont {J.}~\bibnamefont {Lee}},\ }\bibfield  {title} {\bibinfo {title} {Greenberger-horne-zeilinger theorem for $n$ qudits},\ }\href {https://doi.org/10.1103/PhysRevA.88.042101} {\bibfield  {journal} {\bibinfo  {journal} {Phys. Rev. A}\ }\textbf {\bibinfo {volume} {88}},\ \bibinfo {pages} {042101} (\bibinfo {year} {2013})}\BibitemShut {NoStop}%
\bibitem [{\citenamefont {Ryu}\ \emph {et~al.}(2014)\citenamefont {Ryu}, \citenamefont {Lee}, \citenamefont {Yin}, \citenamefont {Rahaman}, \citenamefont {Angelakis}, \citenamefont {Lee},\ and\ \citenamefont {\ifmmode~\dot{Z}\else \.{Z}\fi{}ukowski}}]{ryu2014multisetting}%
  \BibitemOpen
  \bibfield  {author} {\bibinfo {author} {\bibfnamefont {J.}~\bibnamefont {Ryu}}, \bibinfo {author} {\bibfnamefont {C.}~\bibnamefont {Lee}}, \bibinfo {author} {\bibfnamefont {Z.}~\bibnamefont {Yin}}, \bibinfo {author} {\bibfnamefont {R.}~\bibnamefont {Rahaman}}, \bibinfo {author} {\bibfnamefont {D.~G.}\ \bibnamefont {Angelakis}}, \bibinfo {author} {\bibfnamefont {J.}~\bibnamefont {Lee}},\ and\ \bibinfo {author} {\bibfnamefont {M.}~\bibnamefont {\ifmmode~\dot{Z}\else \.{Z}\fi{}ukowski}},\ }\bibfield  {title} {\bibinfo {title} {Multisetting greenberger-horne-zeilinger theorem},\ }\href {https://doi.org/10.1103/PhysRevA.89.024103} {\bibfield  {journal} {\bibinfo  {journal} {Phys. Rev. A}\ }\textbf {\bibinfo {volume} {89}},\ \bibinfo {pages} {024103} (\bibinfo {year} {2014})}\BibitemShut {NoStop}%
\bibitem [{\citenamefont {Lawrence}(2014)}]{lawrence_rotational_2014}%
  \BibitemOpen
  \bibfield  {author} {\bibinfo {author} {\bibfnamefont {J.}~\bibnamefont {Lawrence}},\ }\bibfield  {title} {\bibinfo {title} {Rotational covariance and {Greenberger}-{Horne}-{Zeilinger} theorems for three or more particles of any dimension},\ }\href {https://doi.org/10.1103/PhysRevA.89.012105} {\bibfield  {journal} {\bibinfo  {journal} {Physical Review A}\ }\textbf {\bibinfo {volume} {89}},\ \bibinfo {pages} {012105} (\bibinfo {year} {2014})}\BibitemShut {NoStop}%
\bibitem [{\citenamefont {Hillery}\ \emph {et~al.}(1999)\citenamefont {Hillery}, \citenamefont {Bužek},\ and\ \citenamefont {Berthiaume}}]{hillery_quantum_1999}%
  \BibitemOpen
  \bibfield  {author} {\bibinfo {author} {\bibfnamefont {M.}~\bibnamefont {Hillery}}, \bibinfo {author} {\bibfnamefont {V.}~\bibnamefont {Bužek}},\ and\ \bibinfo {author} {\bibfnamefont {A.}~\bibnamefont {Berthiaume}},\ }\bibfield  {title} {\bibinfo {title} {Quantum secret sharing},\ }\href {https://doi.org/10.1103/PhysRevA.59.1829} {\bibfield  {journal} {\bibinfo  {journal} {Physical Review A}\ }\textbf {\bibinfo {volume} {59}},\ \bibinfo {pages} {1829} (\bibinfo {year} {1999})},\ \bibinfo {note} {publisher: American Physical Society}\BibitemShut {NoStop}%
\bibitem [{\citenamefont {Qin}\ and\ \citenamefont {Tso}(2018)}]{qin2018efficient}%
  \BibitemOpen
  \bibfield  {author} {\bibinfo {author} {\bibfnamefont {H.}~\bibnamefont {Qin}}\ and\ \bibinfo {author} {\bibfnamefont {R.}~\bibnamefont {Tso}},\ }\bibfield  {title} {\bibinfo {title} {Efficient quantum secret sharing based on special multi-dimensional ghz state},\ }\href {https://doi.org/10.1007/s11082-018-1435-y} {\bibfield  {journal} {\bibinfo  {journal} {Optical and Quantum Electronics}\ }\textbf {\bibinfo {volume} {50}},\ \bibinfo {pages} {167} (\bibinfo {year} {2018})}\BibitemShut {NoStop}%
\bibitem [{\citenamefont {Bai}\ \emph {et~al.}(2021)\citenamefont {Bai}, \citenamefont {Zhang},\ and\ \citenamefont {Liu}}]{bai2021verifiable}%
  \BibitemOpen
  \bibfield  {author} {\bibinfo {author} {\bibfnamefont {C.-M.}\ \bibnamefont {Bai}}, \bibinfo {author} {\bibfnamefont {S.}~\bibnamefont {Zhang}},\ and\ \bibinfo {author} {\bibfnamefont {L.}~\bibnamefont {Liu}},\ }\bibfield  {title} {\bibinfo {title} {Verifiable quantum secret sharing scheme using d-dimensional ghz state},\ }\href {https://doi.org/10.1007/s10773-021-04955-1} {\bibfield  {journal} {\bibinfo  {journal} {International Journal of Theoretical Physics}\ }\textbf {\bibinfo {volume} {60}},\ \bibinfo {pages} {3993} (\bibinfo {year} {2021})}\BibitemShut {NoStop}%
\bibitem [{\citenamefont {Hu}\ \emph {et~al.}(2021)\citenamefont {Hu}, \citenamefont {Zhou}, \citenamefont {Li}, \citenamefont {Fan},\ and\ \citenamefont {Tan}}]{hu2021anovel}%
  \BibitemOpen
  \bibfield  {author} {\bibinfo {author} {\bibfnamefont {W.}~\bibnamefont {Hu}}, \bibinfo {author} {\bibfnamefont {R.-G.}\ \bibnamefont {Zhou}}, \bibinfo {author} {\bibfnamefont {X.}~\bibnamefont {Li}}, \bibinfo {author} {\bibfnamefont {P.}~\bibnamefont {Fan}},\ and\ \bibinfo {author} {\bibfnamefont {C.}~\bibnamefont {Tan}},\ }\bibfield  {title} {\bibinfo {title} {A novel dynamic quantum secret sharing in high-dimensional quantum system},\ }\href {https://doi.org/10.1007/s11128-021-03103-2} {\bibfield  {journal} {\bibinfo  {journal} {Quantum Information Processing}\ }\textbf {\bibinfo {volume} {20}},\ \bibinfo {pages} {159} (\bibinfo {year} {2021})}\BibitemShut {NoStop}%
\bibitem [{\citenamefont {Guo}\ \emph {et~al.}(2022)\citenamefont {Guo}, \citenamefont {Li}, \citenamefont {Wei}, \citenamefont {Tang},\ and\ \citenamefont {Cao}}]{guo2022multiparty}%
  \BibitemOpen
  \bibfield  {author} {\bibinfo {author} {\bibfnamefont {H.}~\bibnamefont {Guo}}, \bibinfo {author} {\bibfnamefont {Y.}~\bibnamefont {Li}}, \bibinfo {author} {\bibfnamefont {J.}~\bibnamefont {Wei}}, \bibinfo {author} {\bibfnamefont {J.}~\bibnamefont {Tang}},\ and\ \bibinfo {author} {\bibfnamefont {Y.}~\bibnamefont {Cao}},\ }\bibfield  {title} {\bibinfo {title} {Multi-party deterministic secure quantum communication using d-dimension ghz state},\ }\href {https://doi.org/10.1142/S0217984922501299} {\bibfield  {journal} {\bibinfo  {journal} {Modern Physics Letters B}\ }\textbf {\bibinfo {volume} {36}},\ \bibinfo {pages} {2250129} (\bibinfo {year} {2022})},\ \Eprint {https://arxiv.org/abs/https://doi.org/10.1142/S0217984922501299} {https://doi.org/10.1142/S0217984922501299} \BibitemShut {NoStop}%
\bibitem [{\citenamefont {Erhard}\ \emph {et~al.}(2018)\citenamefont {Erhard}, \citenamefont {Malik}, \citenamefont {Krenn},\ and\ \citenamefont {Zeilinger}}]{erhard2018experimental}%
  \BibitemOpen
  \bibfield  {author} {\bibinfo {author} {\bibfnamefont {M.}~\bibnamefont {Erhard}}, \bibinfo {author} {\bibfnamefont {M.}~\bibnamefont {Malik}}, \bibinfo {author} {\bibfnamefont {M.}~\bibnamefont {Krenn}},\ and\ \bibinfo {author} {\bibfnamefont {A.}~\bibnamefont {Zeilinger}},\ }\bibfield  {title} {\bibinfo {title} {Experimental greenberger--horne--zeilinger entanglement beyond qubits},\ }\href {https://doi.org/10.1038/s41566-018-0257-6} {\bibfield  {journal} {\bibinfo  {journal} {Nature Photonics}\ }\textbf {\bibinfo {volume} {12}},\ \bibinfo {pages} {759} (\bibinfo {year} {2018})}\BibitemShut {NoStop}%
\bibitem [{\citenamefont {Imany}\ \emph {et~al.}(2019)\citenamefont {Imany}, \citenamefont {Jaramillo-Villegas}, \citenamefont {Alshaykh}, \citenamefont {Lukens}, \citenamefont {Odele}, \citenamefont {Moore}, \citenamefont {Leaird}, \citenamefont {Qi},\ and\ \citenamefont {Weiner}}]{imany2019highdimensional}%
  \BibitemOpen
  \bibfield  {author} {\bibinfo {author} {\bibfnamefont {P.}~\bibnamefont {Imany}}, \bibinfo {author} {\bibfnamefont {J.~A.}\ \bibnamefont {Jaramillo-Villegas}}, \bibinfo {author} {\bibfnamefont {M.~S.}\ \bibnamefont {Alshaykh}}, \bibinfo {author} {\bibfnamefont {J.~M.}\ \bibnamefont {Lukens}}, \bibinfo {author} {\bibfnamefont {O.~D.}\ \bibnamefont {Odele}}, \bibinfo {author} {\bibfnamefont {A.~J.}\ \bibnamefont {Moore}}, \bibinfo {author} {\bibfnamefont {D.~E.}\ \bibnamefont {Leaird}}, \bibinfo {author} {\bibfnamefont {M.}~\bibnamefont {Qi}},\ and\ \bibinfo {author} {\bibfnamefont {A.~M.}\ \bibnamefont {Weiner}},\ }\bibfield  {title} {\bibinfo {title} {High-dimensional optical quantum logic in large operational spaces},\ }\href {https://doi.org/10.1038/s41534-019-0173-8} {\bibfield  {journal} {\bibinfo  {journal} {npj Quantum Information}\ }\textbf {\bibinfo {volume} {5}},\ \bibinfo {pages} {59} (\bibinfo {year} {2019})}\BibitemShut {NoStop}%
\bibitem [{\citenamefont {Paesani}\ \emph {et~al.}(2021)\citenamefont {Paesani}, \citenamefont {Bulmer}, \citenamefont {Jones}, \citenamefont {Santagati},\ and\ \citenamefont {Laing}}]{paesani2021scheme}%
  \BibitemOpen
  \bibfield  {author} {\bibinfo {author} {\bibfnamefont {S.}~\bibnamefont {Paesani}}, \bibinfo {author} {\bibfnamefont {J.~F.~F.}\ \bibnamefont {Bulmer}}, \bibinfo {author} {\bibfnamefont {A.~E.}\ \bibnamefont {Jones}}, \bibinfo {author} {\bibfnamefont {R.}~\bibnamefont {Santagati}},\ and\ \bibinfo {author} {\bibfnamefont {A.}~\bibnamefont {Laing}},\ }\bibfield  {title} {\bibinfo {title} {Scheme for universal high-dimensional quantum computation with linear optics},\ }\href {https://doi.org/10.1103/PhysRevLett.126.230504} {\bibfield  {journal} {\bibinfo  {journal} {Phys. Rev. Lett.}\ }\textbf {\bibinfo {volume} {126}},\ \bibinfo {pages} {230504} (\bibinfo {year} {2021})}\BibitemShut {NoStop}%
\bibitem [{\citenamefont {Bell}\ \emph {et~al.}(2022)\citenamefont {Bell}, \citenamefont {Bulmer}, \citenamefont {Jones}, \citenamefont {Paesani}, \citenamefont {McCutcheon},\ and\ \citenamefont {Laing}}]{bell2022protocol}%
  \BibitemOpen
  \bibfield  {author} {\bibinfo {author} {\bibfnamefont {T.~J.}\ \bibnamefont {Bell}}, \bibinfo {author} {\bibfnamefont {J.~F.~F.}\ \bibnamefont {Bulmer}}, \bibinfo {author} {\bibfnamefont {A.~E.}\ \bibnamefont {Jones}}, \bibinfo {author} {\bibfnamefont {S.}~\bibnamefont {Paesani}}, \bibinfo {author} {\bibfnamefont {D.~P.~S.}\ \bibnamefont {McCutcheon}},\ and\ \bibinfo {author} {\bibfnamefont {A.}~\bibnamefont {Laing}},\ }\bibfield  {title} {\bibinfo {title} {Protocol for generation of high-dimensional entanglement from an array of non-interacting photon emitters},\ }\href {https://doi.org/10.1088/1367-2630/ac475d} {\bibfield  {journal} {\bibinfo  {journal} {New Journal of Physics}\ }\textbf {\bibinfo {volume} {24}},\ \bibinfo {pages} {013032} (\bibinfo {year} {2022})}\BibitemShut {NoStop}%
\bibitem [{\citenamefont {Cervera-Lierta}\ \emph {et~al.}(2022)\citenamefont {Cervera-Lierta}, \citenamefont {Krenn}, \citenamefont {Aspuru-Guzik},\ and\ \citenamefont {Galda}}]{lierta2022experimental}%
  \BibitemOpen
  \bibfield  {author} {\bibinfo {author} {\bibfnamefont {A.}~\bibnamefont {Cervera-Lierta}}, \bibinfo {author} {\bibfnamefont {M.}~\bibnamefont {Krenn}}, \bibinfo {author} {\bibfnamefont {A.}~\bibnamefont {Aspuru-Guzik}},\ and\ \bibinfo {author} {\bibfnamefont {A.}~\bibnamefont {Galda}},\ }\bibfield  {title} {\bibinfo {title} {Experimental high-dimensional greenberger-horne-zeilinger entanglement with superconducting transmon qutrits},\ }\href {https://doi.org/10.1103/PhysRevApplied.17.024062} {\bibfield  {journal} {\bibinfo  {journal} {Phys. Rev. Appl.}\ }\textbf {\bibinfo {volume} {17}},\ \bibinfo {pages} {024062} (\bibinfo {year} {2022})}\BibitemShut {NoStop}%
\bibitem [{\citenamefont {Zhao}\ \emph {et~al.}(2024)\citenamefont {Zhao}, \citenamefont {Yang}, \citenamefont {Li},\ and\ \citenamefont {Shao}}]{zhao2024dissipative}%
  \BibitemOpen
  \bibfield  {author} {\bibinfo {author} {\bibfnamefont {Y.}~\bibnamefont {Zhao}}, \bibinfo {author} {\bibfnamefont {Y.-Q.}\ \bibnamefont {Yang}}, \bibinfo {author} {\bibfnamefont {W.}~\bibnamefont {Li}},\ and\ \bibinfo {author} {\bibfnamefont {X.-Q.}\ \bibnamefont {Shao}},\ }\bibfield  {title} {\bibinfo {title} {{Dissipative stabilization of high-dimensional GHZ states for neutral atoms}},\ }\href {https://doi.org/10.1063/5.0192602} {\bibfield  {journal} {\bibinfo  {journal} {Applied Physics Letters}\ }\textbf {\bibinfo {volume} {124}},\ \bibinfo {pages} {114001} (\bibinfo {year} {2024})}\BibitemShut {NoStop}%
\bibitem [{\citenamefont {Zou}\ and\ \citenamefont {Mathis}(2004)}]{zou2004onestep}%
  \BibitemOpen
  \bibfield  {author} {\bibinfo {author} {\bibfnamefont {X.}~\bibnamefont {Zou}}\ and\ \bibinfo {author} {\bibfnamefont {W.}~\bibnamefont {Mathis}},\ }\bibfield  {title} {\bibinfo {title} {One-step implementation of maximally entangled states of many three-level atoms in microwave cavity qed},\ }\href {https://doi.org/10.1103/PhysRevA.70.035802} {\bibfield  {journal} {\bibinfo  {journal} {Phys. Rev. A}\ }\textbf {\bibinfo {volume} {70}},\ \bibinfo {pages} {035802} (\bibinfo {year} {2004})}\BibitemShut {NoStop}%
\bibitem [{\citenamefont {Kitagawa}\ and\ \citenamefont {Ueda}(1993)}]{kitagawa1993squeezed}%
  \BibitemOpen
  \bibfield  {author} {\bibinfo {author} {\bibfnamefont {M.}~\bibnamefont {Kitagawa}}\ and\ \bibinfo {author} {\bibfnamefont {M.}~\bibnamefont {Ueda}},\ }\bibfield  {title} {\bibinfo {title} {Squeezed spin states},\ }\href {https://doi.org/10.1103/PhysRevA.47.5138} {\bibfield  {journal} {\bibinfo  {journal} {Phys. Rev. A}\ }\textbf {\bibinfo {volume} {47}},\ \bibinfo {pages} {5138} (\bibinfo {year} {1993})}\BibitemShut {NoStop}%
\bibitem [{\citenamefont {Pezz\`e}\ \emph {et~al.}(2018)\citenamefont {Pezz\`e}, \citenamefont {Smerzi}, \citenamefont {Oberthaler}, \citenamefont {Schmied},\ and\ \citenamefont {Treutlein}}]{pezze2018RMP}%
  \BibitemOpen
  \bibfield  {author} {\bibinfo {author} {\bibfnamefont {L.}~\bibnamefont {Pezz\`e}}, \bibinfo {author} {\bibfnamefont {A.}~\bibnamefont {Smerzi}}, \bibinfo {author} {\bibfnamefont {M.~K.}\ \bibnamefont {Oberthaler}}, \bibinfo {author} {\bibfnamefont {R.}~\bibnamefont {Schmied}},\ and\ \bibinfo {author} {\bibfnamefont {P.}~\bibnamefont {Treutlein}},\ }\bibfield  {title} {\bibinfo {title} {Quantum metrology with nonclassical states of atomic ensembles},\ }\href {https://doi.org/10.1103/RevModPhys.90.035005} {\bibfield  {journal} {\bibinfo  {journal} {Rev. Mod. Phys.}\ }\textbf {\bibinfo {volume} {90}},\ \bibinfo {pages} {035005} (\bibinfo {year} {2018})}\BibitemShut {NoStop}%
\bibitem [{\citenamefont {Agarwal}\ \emph {et~al.}(1997)\citenamefont {Agarwal}, \citenamefont {Puri},\ and\ \citenamefont {Singh}}]{agarwal1997atomic}%
  \BibitemOpen
  \bibfield  {author} {\bibinfo {author} {\bibfnamefont {G.}~\bibnamefont {Agarwal}}, \bibinfo {author} {\bibfnamefont {R.}~\bibnamefont {Puri}},\ and\ \bibinfo {author} {\bibfnamefont {R.}~\bibnamefont {Singh}},\ }\bibfield  {title} {\bibinfo {title} {Atomic {Schrödinger} cat states},\ }\href {https://doi.org/10.1103/PhysRevA.56.2249} {\bibfield  {journal} {\bibinfo  {journal} {Physical Review A}\ }\textbf {\bibinfo {volume} {56}},\ \bibinfo {pages} {2249} (\bibinfo {year} {1997})}\BibitemShut {NoStop}%
\bibitem [{\citenamefont {Low}\ \emph {et~al.}(2020)\citenamefont {Low}, \citenamefont {White}, \citenamefont {Cox}, \citenamefont {Day},\ and\ \citenamefont {Senko}}]{low2020practical}%
  \BibitemOpen
  \bibfield  {author} {\bibinfo {author} {\bibfnamefont {P.~J.}\ \bibnamefont {Low}}, \bibinfo {author} {\bibfnamefont {B.~M.}\ \bibnamefont {White}}, \bibinfo {author} {\bibfnamefont {A.~A.}\ \bibnamefont {Cox}}, \bibinfo {author} {\bibfnamefont {M.~L.}\ \bibnamefont {Day}},\ and\ \bibinfo {author} {\bibfnamefont {C.}~\bibnamefont {Senko}},\ }\bibfield  {title} {\bibinfo {title} {Practical trapped-ion protocols for universal qudit-based quantum computing},\ }\href {https://doi.org/10.1103/PhysRevResearch.2.033128} {\bibfield  {journal} {\bibinfo  {journal} {Phys. Rev. Res.}\ }\textbf {\bibinfo {volume} {2}},\ \bibinfo {pages} {033128} (\bibinfo {year} {2020})}\BibitemShut {NoStop}%
\bibitem [{\citenamefont {Ringbauer}\ \emph {et~al.}(2022)\citenamefont {Ringbauer}, \citenamefont {Meth}, \citenamefont {Postler}, \citenamefont {Stricker}, \citenamefont {Blatt}, \citenamefont {Schindler},\ and\ \citenamefont {Monz}}]{ringbauer2022auniversal}%
  \BibitemOpen
  \bibfield  {author} {\bibinfo {author} {\bibfnamefont {M.}~\bibnamefont {Ringbauer}}, \bibinfo {author} {\bibfnamefont {M.}~\bibnamefont {Meth}}, \bibinfo {author} {\bibfnamefont {L.}~\bibnamefont {Postler}}, \bibinfo {author} {\bibfnamefont {R.}~\bibnamefont {Stricker}}, \bibinfo {author} {\bibfnamefont {R.}~\bibnamefont {Blatt}}, \bibinfo {author} {\bibfnamefont {P.}~\bibnamefont {Schindler}},\ and\ \bibinfo {author} {\bibfnamefont {T.}~\bibnamefont {Monz}},\ }\bibfield  {title} {\bibinfo {title} {A universal qudit quantum processor with trapped ions},\ }\href {https://doi.org/10.1038/s41567-022-01658-0} {\bibfield  {journal} {\bibinfo  {journal} {Nature Physics}\ }\textbf {\bibinfo {volume} {18}},\ \bibinfo {pages} {1053} (\bibinfo {year} {2022})}\BibitemShut {NoStop}%
\bibitem [{\citenamefont {Hrmo}\ \emph {et~al.}(2023)\citenamefont {Hrmo}, \citenamefont {Wilhelm}, \citenamefont {Gerster}, \citenamefont {van Mourik}, \citenamefont {Huber}, \citenamefont {Blatt}, \citenamefont {Schindler}, \citenamefont {Monz},\ and\ \citenamefont {Ringbauer}}]{hrmo2023native}%
  \BibitemOpen
  \bibfield  {author} {\bibinfo {author} {\bibfnamefont {P.}~\bibnamefont {Hrmo}}, \bibinfo {author} {\bibfnamefont {B.}~\bibnamefont {Wilhelm}}, \bibinfo {author} {\bibfnamefont {L.}~\bibnamefont {Gerster}}, \bibinfo {author} {\bibfnamefont {M.~W.}\ \bibnamefont {van Mourik}}, \bibinfo {author} {\bibfnamefont {M.}~\bibnamefont {Huber}}, \bibinfo {author} {\bibfnamefont {R.}~\bibnamefont {Blatt}}, \bibinfo {author} {\bibfnamefont {P.}~\bibnamefont {Schindler}}, \bibinfo {author} {\bibfnamefont {T.}~\bibnamefont {Monz}},\ and\ \bibinfo {author} {\bibfnamefont {M.}~\bibnamefont {Ringbauer}},\ }\bibfield  {title} {\bibinfo {title} {Native qudit entanglement in a trapped ion quantum processor},\ }\href {https://doi.org/10.1038/s41467-023-37375-2} {\bibfield  {journal} {\bibinfo  {journal} {Nature Communications}\ }\textbf {\bibinfo {volume} {14}},\ \bibinfo {pages} {2242} (\bibinfo {year} {2023})}\BibitemShut {NoStop}%
\bibitem [{\citenamefont {Kaewuam}\ \emph {et~al.}(2018)\citenamefont {Kaewuam}, \citenamefont {Roy}, \citenamefont {Tan}, \citenamefont {Arnold},\ and\ \citenamefont {Barrett}}]{kaewuam2019spectroscopy}%
  \BibitemOpen
  \bibfield  {author} {\bibinfo {author} {\bibfnamefont {R.}~\bibnamefont {Kaewuam}}, \bibinfo {author} {\bibfnamefont {A.}~\bibnamefont {Roy}}, \bibinfo {author} {\bibfnamefont {T.~R.}\ \bibnamefont {Tan}}, \bibinfo {author} {\bibfnamefont {K.~J.}\ \bibnamefont {Arnold}},\ and\ \bibinfo {author} {\bibfnamefont {M.~D.}\ \bibnamefont {Barrett}},\ }\bibfield  {title} {\bibinfo {title} {Laser spectroscopy of 176lu+},\ }\href {https://doi.org/10.1080/09500340.2017.1411539} {\bibfield  {journal} {\bibinfo  {journal} {Journal of Modern Optics}\ }\textbf {\bibinfo {volume} {65}},\ \bibinfo {pages} {592} (\bibinfo {year} {2018})},\ \Eprint {https://arxiv.org/abs/https://doi.org/10.1080/09500340.2017.1411539} {https://doi.org/10.1080/09500340.2017.1411539} \BibitemShut {NoStop}%
\bibitem [{\citenamefont {Allcock}\ \emph {et~al.}(2021)\citenamefont {Allcock}, \citenamefont {Campbell}, \citenamefont {Chiaverini}, \citenamefont {Chuang}, \citenamefont {Hudson}, \citenamefont {Moore}, \citenamefont {Ransford}, \citenamefont {Roman}, \citenamefont {Sage},\ and\ \citenamefont {Wineland}}]{allcock2021omg}%
  \BibitemOpen
  \bibfield  {author} {\bibinfo {author} {\bibfnamefont {D.~T.~C.}\ \bibnamefont {Allcock}}, \bibinfo {author} {\bibfnamefont {W.~C.}\ \bibnamefont {Campbell}}, \bibinfo {author} {\bibfnamefont {J.}~\bibnamefont {Chiaverini}}, \bibinfo {author} {\bibfnamefont {I.~L.}\ \bibnamefont {Chuang}}, \bibinfo {author} {\bibfnamefont {E.~R.}\ \bibnamefont {Hudson}}, \bibinfo {author} {\bibfnamefont {I.~D.}\ \bibnamefont {Moore}}, \bibinfo {author} {\bibfnamefont {A.}~\bibnamefont {Ransford}}, \bibinfo {author} {\bibfnamefont {C.}~\bibnamefont {Roman}}, \bibinfo {author} {\bibfnamefont {J.~M.}\ \bibnamefont {Sage}},\ and\ \bibinfo {author} {\bibfnamefont {D.~J.}\ \bibnamefont {Wineland}},\ }\bibfield  {title} {\bibinfo {title} {{omg blueprint for trapped ion quantum computing with metastable states}},\ }\href {https://doi.org/10.1063/5.0069544} {\bibfield  {journal} {\bibinfo  {journal} {Applied Physics Letters}\ }\textbf {\bibinfo {volume} {119}},\ \bibinfo {pages} {214002} (\bibinfo {year} {2021})}\BibitemShut
  {NoStop}%
\bibitem [{\citenamefont {Malik}\ \emph {et~al.}(2016)\citenamefont {Malik}, \citenamefont {Erhard}, \citenamefont {Huber}, \citenamefont {Krenn}, \citenamefont {Fickler},\ and\ \citenamefont {Zeilinger}}]{malik2016multiphoton}%
  \BibitemOpen
  \bibfield  {author} {\bibinfo {author} {\bibfnamefont {M.}~\bibnamefont {Malik}}, \bibinfo {author} {\bibfnamefont {M.}~\bibnamefont {Erhard}}, \bibinfo {author} {\bibfnamefont {M.}~\bibnamefont {Huber}}, \bibinfo {author} {\bibfnamefont {M.}~\bibnamefont {Krenn}}, \bibinfo {author} {\bibfnamefont {R.}~\bibnamefont {Fickler}},\ and\ \bibinfo {author} {\bibfnamefont {A.}~\bibnamefont {Zeilinger}},\ }\bibfield  {title} {\bibinfo {title} {Multi-photon entanglement in high dimensions},\ }\href {https://doi.org/10.1038/nphoton.2016.12} {\bibfield  {journal} {\bibinfo  {journal} {Nature Photonics}\ }\textbf {\bibinfo {volume} {10}},\ \bibinfo {pages} {248} (\bibinfo {year} {2016})}\BibitemShut {NoStop}%
\bibitem [{\citenamefont {Fickler}\ \emph {et~al.}(2014)\citenamefont {Fickler}, \citenamefont {Lapkiewicz}, \citenamefont {Huber}, \citenamefont {Lavery}, \citenamefont {Padgett},\ and\ \citenamefont {Zeilinger}}]{fickler2014interface}%
  \BibitemOpen
  \bibfield  {author} {\bibinfo {author} {\bibfnamefont {R.}~\bibnamefont {Fickler}}, \bibinfo {author} {\bibfnamefont {R.}~\bibnamefont {Lapkiewicz}}, \bibinfo {author} {\bibfnamefont {M.}~\bibnamefont {Huber}}, \bibinfo {author} {\bibfnamefont {M.~P.}\ \bibnamefont {Lavery}}, \bibinfo {author} {\bibfnamefont {M.~J.}\ \bibnamefont {Padgett}},\ and\ \bibinfo {author} {\bibfnamefont {A.}~\bibnamefont {Zeilinger}},\ }\bibfield  {title} {\bibinfo {title} {Interface between path and orbital angular momentum entanglement for high-dimensional photonic quantum information},\ }\href {https://doi.org/10.1038/ncomms5502} {\bibfield  {journal} {\bibinfo  {journal} {Nature Communications}\ }\textbf {\bibinfo {volume} {5}},\ \bibinfo {pages} {4502} (\bibinfo {year} {2014})}\BibitemShut {NoStop}%
\bibitem [{\citenamefont {G{\"a}rttner}\ \emph {et~al.}(2017)\citenamefont {G{\"a}rttner}, \citenamefont {Bohnet}, \citenamefont {Safavi-Naini}, \citenamefont {Wall}, \citenamefont {Bollinger},\ and\ \citenamefont {Rey}}]{gärttner2017measuring}%
  \BibitemOpen
  \bibfield  {author} {\bibinfo {author} {\bibfnamefont {M.}~\bibnamefont {G{\"a}rttner}}, \bibinfo {author} {\bibfnamefont {J.~G.}\ \bibnamefont {Bohnet}}, \bibinfo {author} {\bibfnamefont {A.}~\bibnamefont {Safavi-Naini}}, \bibinfo {author} {\bibfnamefont {M.~L.}\ \bibnamefont {Wall}}, \bibinfo {author} {\bibfnamefont {J.~J.}\ \bibnamefont {Bollinger}},\ and\ \bibinfo {author} {\bibfnamefont {A.~M.}\ \bibnamefont {Rey}},\ }\bibfield  {title} {\bibinfo {title} {Measuring out-of-time-order correlations and multiple quantum spectra in a trapped-ion quantum magnet},\ }\href {https://doi.org/10.1038/nphys4119} {\bibfield  {journal} {\bibinfo  {journal} {Nature Physics}\ }\textbf {\bibinfo {volume} {13}},\ \bibinfo {pages} {781} (\bibinfo {year} {2017})}\BibitemShut {NoStop}%
\bibitem [{\citenamefont {G\"arttner}\ \emph {et~al.}(2018)\citenamefont {G\"arttner}, \citenamefont {Hauke},\ and\ \citenamefont {Rey}}]{garttner2018relating}%
  \BibitemOpen
  \bibfield  {author} {\bibinfo {author} {\bibfnamefont {M.}~\bibnamefont {G\"arttner}}, \bibinfo {author} {\bibfnamefont {P.}~\bibnamefont {Hauke}},\ and\ \bibinfo {author} {\bibfnamefont {A.~M.}\ \bibnamefont {Rey}},\ }\bibfield  {title} {\bibinfo {title} {Relating out-of-time-order correlations to entanglement via multiple-quantum coherences},\ }\href {https://doi.org/10.1103/PhysRevLett.120.040402} {\bibfield  {journal} {\bibinfo  {journal} {Phys. Rev. Lett.}\ }\textbf {\bibinfo {volume} {120}},\ \bibinfo {pages} {040402} (\bibinfo {year} {2018})}\BibitemShut {NoStop}%
\bibitem [{\citenamefont {Lewis-Swan}\ \emph {et~al.}(2019)\citenamefont {Lewis-Swan}, \citenamefont {Safavi-Naini}, \citenamefont {Bollinger},\ and\ \citenamefont {Rey}}]{lewisswan2019unifying}%
  \BibitemOpen
  \bibfield  {author} {\bibinfo {author} {\bibfnamefont {R.~J.}\ \bibnamefont {Lewis-Swan}}, \bibinfo {author} {\bibfnamefont {A.}~\bibnamefont {Safavi-Naini}}, \bibinfo {author} {\bibfnamefont {J.~J.}\ \bibnamefont {Bollinger}},\ and\ \bibinfo {author} {\bibfnamefont {A.~M.}\ \bibnamefont {Rey}},\ }\bibfield  {title} {\bibinfo {title} {Unifying scrambling, thermalization and entanglement through measurement of fidelity out-of-time-order correlators in the dicke model},\ }\href {https://doi.org/10.1038/s41467-019-09436-y} {\bibfield  {journal} {\bibinfo  {journal} {Nature Communications}\ }\textbf {\bibinfo {volume} {10}},\ \bibinfo {pages} {1581} (\bibinfo {year} {2019})}\BibitemShut {NoStop}%
\bibitem [{\citenamefont {Colombo}\ \emph {et~al.}(2022)\citenamefont {Colombo}, \citenamefont {Pedrozo-Pe{\~{n}}afiel}, \citenamefont {Adiyatullin}, \citenamefont {Li}, \citenamefont {Mendez}, \citenamefont {Shu},\ and\ \citenamefont {Vuleti{\'{c}}}}]{colombo2022timereversal}%
  \BibitemOpen
  \bibfield  {author} {\bibinfo {author} {\bibfnamefont {S.}~\bibnamefont {Colombo}}, \bibinfo {author} {\bibfnamefont {E.}~\bibnamefont {Pedrozo-Pe{\~{n}}afiel}}, \bibinfo {author} {\bibfnamefont {A.~F.}\ \bibnamefont {Adiyatullin}}, \bibinfo {author} {\bibfnamefont {Z.}~\bibnamefont {Li}}, \bibinfo {author} {\bibfnamefont {E.}~\bibnamefont {Mendez}}, \bibinfo {author} {\bibfnamefont {C.}~\bibnamefont {Shu}},\ and\ \bibinfo {author} {\bibfnamefont {V.}~\bibnamefont {Vuleti{\'{c}}}},\ }\bibfield  {title} {\bibinfo {title} {Time-reversal-based quantum metrology with many-body entangled states},\ }\href {https://doi.org/10.1038/s41567-022-01653-5} {\bibfield  {journal} {\bibinfo  {journal} {Nature Physics}\ }\textbf {\bibinfo {volume} {18}},\ \bibinfo {pages} {925} (\bibinfo {year} {2022})}\BibitemShut {NoStop}%
\bibitem [{\citenamefont {Arnold}\ \emph {et~al.}(2016)\citenamefont {Arnold}, \citenamefont {Kaewuam}, \citenamefont {Roy}, \citenamefont {Paez}, \citenamefont {Wang},\ and\ \citenamefont {Barrett}}]{arnold2016observation}%
  \BibitemOpen
  \bibfield  {author} {\bibinfo {author} {\bibfnamefont {K.~J.}\ \bibnamefont {Arnold}}, \bibinfo {author} {\bibfnamefont {R.}~\bibnamefont {Kaewuam}}, \bibinfo {author} {\bibfnamefont {A.}~\bibnamefont {Roy}}, \bibinfo {author} {\bibfnamefont {E.}~\bibnamefont {Paez}}, \bibinfo {author} {\bibfnamefont {S.}~\bibnamefont {Wang}},\ and\ \bibinfo {author} {\bibfnamefont {M.~D.}\ \bibnamefont {Barrett}},\ }\bibfield  {title} {\bibinfo {title} {Observation of the ${}^{1}{S}_{0}$ to ${}^{3}{D}_{1}$ clock transition in ${}^{175}{\mathrm{lu}}^{+}$},\ }\href {https://doi.org/10.1103/PhysRevA.94.052512} {\bibfield  {journal} {\bibinfo  {journal} {Phys. Rev. A}\ }\textbf {\bibinfo {volume} {94}},\ \bibinfo {pages} {052512} (\bibinfo {year} {2016})}\BibitemShut {NoStop}%
\bibitem [{\citenamefont {M\o{}lmer}\ and\ \citenamefont {S\o{}rensen}(1999)}]{molmer1999multiparticle}%
  \BibitemOpen
  \bibfield  {author} {\bibinfo {author} {\bibfnamefont {K.}~\bibnamefont {M\o{}lmer}}\ and\ \bibinfo {author} {\bibfnamefont {A.}~\bibnamefont {S\o{}rensen}},\ }\bibfield  {title} {\bibinfo {title} {Multiparticle entanglement of hot trapped ions},\ }\href {https://doi.org/10.1103/PhysRevLett.82.1835} {\bibfield  {journal} {\bibinfo  {journal} {Phys. Rev. Lett.}\ }\textbf {\bibinfo {volume} {82}},\ \bibinfo {pages} {1835} (\bibinfo {year} {1999})}\BibitemShut {NoStop}%
\bibitem [{\citenamefont {Benhelm}\ \emph {et~al.}(2008)\citenamefont {Benhelm}, \citenamefont {Kirchmair}, \citenamefont {Roos},\ and\ \citenamefont {Blatt}}]{benhelm2008experimental}%
  \BibitemOpen
  \bibfield  {author} {\bibinfo {author} {\bibfnamefont {J.}~\bibnamefont {Benhelm}}, \bibinfo {author} {\bibfnamefont {G.}~\bibnamefont {Kirchmair}}, \bibinfo {author} {\bibfnamefont {C.~F.}\ \bibnamefont {Roos}},\ and\ \bibinfo {author} {\bibfnamefont {R.}~\bibnamefont {Blatt}},\ }\bibfield  {title} {\bibinfo {title} {Experimental quantum-information processing with ${^{43}\text{C}\text{a}}^{+}$ ions},\ }\href {https://doi.org/10.1103/PhysRevA.77.062306} {\bibfield  {journal} {\bibinfo  {journal} {Phys. Rev. A}\ }\textbf {\bibinfo {volume} {77}},\ \bibinfo {pages} {062306} (\bibinfo {year} {2008})}\BibitemShut {NoStop}%
\bibitem [{\citenamefont {Roberts}\ \emph {et~al.}(2000)\citenamefont {Roberts}, \citenamefont {Taylor}, \citenamefont {Barwood}, \citenamefont {Rowley},\ and\ \citenamefont {Gill}}]{roberts2000observation}%
  \BibitemOpen
  \bibfield  {author} {\bibinfo {author} {\bibfnamefont {M.}~\bibnamefont {Roberts}}, \bibinfo {author} {\bibfnamefont {P.}~\bibnamefont {Taylor}}, \bibinfo {author} {\bibfnamefont {G.~P.}\ \bibnamefont {Barwood}}, \bibinfo {author} {\bibfnamefont {W.~R.~C.}\ \bibnamefont {Rowley}},\ and\ \bibinfo {author} {\bibfnamefont {P.}~\bibnamefont {Gill}},\ }\bibfield  {title} {\bibinfo {title} {Observation of the ${}^{2}{S}_{1/2}{\ensuremath{-}}^{2}{F}_{7/2}$ electric octupole transition in a single ${}^{171}{\mathrm{yb}}^{+}$ ion},\ }\href {https://doi.org/10.1103/PhysRevA.62.020501} {\bibfield  {journal} {\bibinfo  {journal} {Phys. Rev. A}\ }\textbf {\bibinfo {volume} {62}},\ \bibinfo {pages} {020501} (\bibinfo {year} {2000})}\BibitemShut {NoStop}%
\bibitem [{\citenamefont {Sackett}\ \emph {et~al.}(2000)\citenamefont {Sackett}, \citenamefont {Kielpinski}, \citenamefont {King}, \citenamefont {Langer}, \citenamefont {Meyer}, \citenamefont {Myatt}, \citenamefont {Rowe}, \citenamefont {Turchette}, \citenamefont {Itano}, \citenamefont {Wineland},\ and\ \citenamefont {Monroe}}]{sackett2000experimental}%
  \BibitemOpen
  \bibfield  {author} {\bibinfo {author} {\bibfnamefont {C.~A.}\ \bibnamefont {Sackett}}, \bibinfo {author} {\bibfnamefont {D.}~\bibnamefont {Kielpinski}}, \bibinfo {author} {\bibfnamefont {B.~E.}\ \bibnamefont {King}}, \bibinfo {author} {\bibfnamefont {C.}~\bibnamefont {Langer}}, \bibinfo {author} {\bibfnamefont {V.}~\bibnamefont {Meyer}}, \bibinfo {author} {\bibfnamefont {C.~J.}\ \bibnamefont {Myatt}}, \bibinfo {author} {\bibfnamefont {M.}~\bibnamefont {Rowe}}, \bibinfo {author} {\bibfnamefont {Q.~A.}\ \bibnamefont {Turchette}}, \bibinfo {author} {\bibfnamefont {W.~M.}\ \bibnamefont {Itano}}, \bibinfo {author} {\bibfnamefont {D.~J.}\ \bibnamefont {Wineland}},\ and\ \bibinfo {author} {\bibfnamefont {C.}~\bibnamefont {Monroe}},\ }\bibfield  {title} {\bibinfo {title} {Experimental entanglement of four particles},\ }\href {https://doi.org/10.1038/35005011} {\bibfield  {journal} {\bibinfo  {journal} {Nature}\ }\textbf {\bibinfo {volume} {404}},\ \bibinfo {pages} {256} (\bibinfo {year} {2000})}\BibitemShut
  {NoStop}%
\bibitem [{\citenamefont {Cadney}\ \emph {et~al.}(2014)\citenamefont {Cadney}, \citenamefont {Huber}, \citenamefont {Linden},\ and\ \citenamefont {Winter}}]{cadney2014inequalities}%
  \BibitemOpen
  \bibfield  {author} {\bibinfo {author} {\bibfnamefont {J.}~\bibnamefont {Cadney}}, \bibinfo {author} {\bibfnamefont {M.}~\bibnamefont {Huber}}, \bibinfo {author} {\bibfnamefont {N.}~\bibnamefont {Linden}},\ and\ \bibinfo {author} {\bibfnamefont {A.}~\bibnamefont {Winter}},\ }\bibfield  {title} {\bibinfo {title} {Inequalities for the ranks of multipartite quantum states},\ }\href {https://doi.org/https://doi.org/10.1016/j.laa.2014.03.035} {\bibfield  {journal} {\bibinfo  {journal} {Linear Algebra and its Applications}\ }\textbf {\bibinfo {volume} {452}},\ \bibinfo {pages} {153} (\bibinfo {year} {2014})}\BibitemShut {NoStop}%
\bibitem [{\citenamefont {Hu}\ \emph {et~al.}(2017)\citenamefont {Hu}, \citenamefont {Chen}, \citenamefont {Vendeiro}, \citenamefont {Urvoy}, \citenamefont {Braverman},\ and\ \citenamefont {Vuleti\ifmmode~\acute{c}\else \'{c}\fi{}}}]{hu2017vacuum}%
  \BibitemOpen
  \bibfield  {author} {\bibinfo {author} {\bibfnamefont {J.}~\bibnamefont {Hu}}, \bibinfo {author} {\bibfnamefont {W.}~\bibnamefont {Chen}}, \bibinfo {author} {\bibfnamefont {Z.}~\bibnamefont {Vendeiro}}, \bibinfo {author} {\bibfnamefont {A.}~\bibnamefont {Urvoy}}, \bibinfo {author} {\bibfnamefont {B.}~\bibnamefont {Braverman}},\ and\ \bibinfo {author} {\bibfnamefont {V.}~\bibnamefont {Vuleti\ifmmode~\acute{c}\else \'{c}\fi{}}},\ }\bibfield  {title} {\bibinfo {title} {Vacuum spin squeezing},\ }\href {https://doi.org/10.1103/PhysRevA.96.050301} {\bibfield  {journal} {\bibinfo  {journal} {Phys. Rev. A}\ }\textbf {\bibinfo {volume} {96}},\ \bibinfo {pages} {050301} (\bibinfo {year} {2017})}\BibitemShut {NoStop}%
\bibitem [{\citenamefont {Norcia}\ \emph {et~al.}(2018)\citenamefont {Norcia}, \citenamefont {Lewis-Swan}, \citenamefont {Cline}, \citenamefont {Zhu}, \citenamefont {Rey},\ and\ \citenamefont {Thompson}}]{norcia2018cavity}%
  \BibitemOpen
  \bibfield  {author} {\bibinfo {author} {\bibfnamefont {M.~A.}\ \bibnamefont {Norcia}}, \bibinfo {author} {\bibfnamefont {R.~J.}\ \bibnamefont {Lewis-Swan}}, \bibinfo {author} {\bibfnamefont {J.~R.~K.}\ \bibnamefont {Cline}}, \bibinfo {author} {\bibfnamefont {B.}~\bibnamefont {Zhu}}, \bibinfo {author} {\bibfnamefont {A.~M.}\ \bibnamefont {Rey}},\ and\ \bibinfo {author} {\bibfnamefont {J.~K.}\ \bibnamefont {Thompson}},\ }\bibfield  {title} {\bibinfo {title} {Cavity-mediated collective spin-exchange interactions in a strontium superradiant laser},\ }\href {https://doi.org/10.1126/science.aar3102} {\bibfield  {journal} {\bibinfo  {journal} {Science}\ }\textbf {\bibinfo {volume} {361}},\ \bibinfo {pages} {259} (\bibinfo {year} {2018})},\ \Eprint {https://arxiv.org/abs/https://www.science.org/doi/pdf/10.1126/science.aar3102} {https://www.science.org/doi/pdf/10.1126/science.aar3102} \BibitemShut {NoStop}%
\bibitem [{\citenamefont {Xu}\ \emph {et~al.}(2022)\citenamefont {Xu}, \citenamefont {Zhang}, \citenamefont {Sun}, \citenamefont {Li}, \citenamefont {Song}, \citenamefont {Xiang}, \citenamefont {Huang}, \citenamefont {Li}, \citenamefont {Shi}, \citenamefont {Chen}, \citenamefont {Song}, \citenamefont {Zheng}, \citenamefont {Nori}, \citenamefont {Wang},\ and\ \citenamefont {Fan}}]{xu2022metrological}%
  \BibitemOpen
  \bibfield  {author} {\bibinfo {author} {\bibfnamefont {K.}~\bibnamefont {Xu}}, \bibinfo {author} {\bibfnamefont {Y.-R.}\ \bibnamefont {Zhang}}, \bibinfo {author} {\bibfnamefont {Z.-H.}\ \bibnamefont {Sun}}, \bibinfo {author} {\bibfnamefont {H.}~\bibnamefont {Li}}, \bibinfo {author} {\bibfnamefont {P.}~\bibnamefont {Song}}, \bibinfo {author} {\bibfnamefont {Z.}~\bibnamefont {Xiang}}, \bibinfo {author} {\bibfnamefont {K.}~\bibnamefont {Huang}}, \bibinfo {author} {\bibfnamefont {H.}~\bibnamefont {Li}}, \bibinfo {author} {\bibfnamefont {Y.-H.}\ \bibnamefont {Shi}}, \bibinfo {author} {\bibfnamefont {C.-T.}\ \bibnamefont {Chen}}, \bibinfo {author} {\bibfnamefont {X.}~\bibnamefont {Song}}, \bibinfo {author} {\bibfnamefont {D.}~\bibnamefont {Zheng}}, \bibinfo {author} {\bibfnamefont {F.}~\bibnamefont {Nori}}, \bibinfo {author} {\bibfnamefont {H.}~\bibnamefont {Wang}},\ and\ \bibinfo {author} {\bibfnamefont {H.}~\bibnamefont {Fan}},\ }\bibfield  {title} {\bibinfo {title} {Metrological characterization of
  non-gaussian entangled states of superconducting qubits},\ }\href {https://doi.org/10.1103/PhysRevLett.128.150501} {\bibfield  {journal} {\bibinfo  {journal} {Phys. Rev. Lett.}\ }\textbf {\bibinfo {volume} {128}},\ \bibinfo {pages} {150501} (\bibinfo {year} {2022})}\BibitemShut {NoStop}%
\bibitem [{\citenamefont {Hazra}\ \emph {et~al.}(2021)\citenamefont {Hazra}, \citenamefont {Bhattacharjee}, \citenamefont {Chand}, \citenamefont {Salunkhe}, \citenamefont {Gopalakrishnan}, \citenamefont {Patankar},\ and\ \citenamefont {Vijay}}]{hazra2021ringresonator}%
  \BibitemOpen
  \bibfield  {author} {\bibinfo {author} {\bibfnamefont {S.}~\bibnamefont {Hazra}}, \bibinfo {author} {\bibfnamefont {A.}~\bibnamefont {Bhattacharjee}}, \bibinfo {author} {\bibfnamefont {M.}~\bibnamefont {Chand}}, \bibinfo {author} {\bibfnamefont {K.~V.}\ \bibnamefont {Salunkhe}}, \bibinfo {author} {\bibfnamefont {S.}~\bibnamefont {Gopalakrishnan}}, \bibinfo {author} {\bibfnamefont {M.~P.}\ \bibnamefont {Patankar}},\ and\ \bibinfo {author} {\bibfnamefont {R.}~\bibnamefont {Vijay}},\ }\bibfield  {title} {\bibinfo {title} {Ring-resonator-based coupling architecture for enhanced connectivity in a superconducting multiqubit network},\ }\href {https://doi.org/10.1103/PhysRevApplied.16.024018} {\bibfield  {journal} {\bibinfo  {journal} {Phys. Rev. Appl.}\ }\textbf {\bibinfo {volume} {16}},\ \bibinfo {pages} {024018} (\bibinfo {year} {2021})}\BibitemShut {NoStop}%
\bibitem [{\citenamefont {Cao}\ \emph {et~al.}(2024)\citenamefont {Cao}, \citenamefont {Eckner}, \citenamefont {Yelin}, \citenamefont {Young}, \citenamefont {Jandura}, \citenamefont {Yan}, \citenamefont {Kim}, \citenamefont {Pupillo}, \citenamefont {Ye}, \citenamefont {Oppong},\ and\ \citenamefont {Kaufman}}]{cao2024multiqubitgates}%
  \BibitemOpen
  \bibfield  {author} {\bibinfo {author} {\bibfnamefont {A.}~\bibnamefont {Cao}}, \bibinfo {author} {\bibfnamefont {W.~J.}\ \bibnamefont {Eckner}}, \bibinfo {author} {\bibfnamefont {T.~L.}\ \bibnamefont {Yelin}}, \bibinfo {author} {\bibfnamefont {A.~W.}\ \bibnamefont {Young}}, \bibinfo {author} {\bibfnamefont {S.}~\bibnamefont {Jandura}}, \bibinfo {author} {\bibfnamefont {L.}~\bibnamefont {Yan}}, \bibinfo {author} {\bibfnamefont {K.}~\bibnamefont {Kim}}, \bibinfo {author} {\bibfnamefont {G.}~\bibnamefont {Pupillo}}, \bibinfo {author} {\bibfnamefont {J.}~\bibnamefont {Ye}}, \bibinfo {author} {\bibfnamefont {N.~D.}\ \bibnamefont {Oppong}},\ and\ \bibinfo {author} {\bibfnamefont {A.~M.}\ \bibnamefont {Kaufman}},\ }\href {https://arxiv.org/abs/2402.16289} {\bibinfo {title} {Multi-qubit gates and schr\"odinger cat states in an optical clock}} (\bibinfo {year} {2024}),\ \Eprint {https://arxiv.org/abs/2402.16289} {arXiv:2402.16289 [quant-ph]} \BibitemShut {NoStop}%
\bibitem [{\citenamefont {Kaubruegger}\ \emph {et~al.}(2021)\citenamefont {Kaubruegger}, \citenamefont {Vasilyev}, \citenamefont {Schulte}, \citenamefont {Hammerer},\ and\ \citenamefont {Zoller}}]{kaubruegger2021quantum}%
  \BibitemOpen
  \bibfield  {author} {\bibinfo {author} {\bibfnamefont {R.}~\bibnamefont {Kaubruegger}}, \bibinfo {author} {\bibfnamefont {D.~V.}\ \bibnamefont {Vasilyev}}, \bibinfo {author} {\bibfnamefont {M.}~\bibnamefont {Schulte}}, \bibinfo {author} {\bibfnamefont {K.}~\bibnamefont {Hammerer}},\ and\ \bibinfo {author} {\bibfnamefont {P.}~\bibnamefont {Zoller}},\ }\bibfield  {title} {\bibinfo {title} {Quantum variational optimization of ramsey interferometry and atomic clocks},\ }\href {https://doi.org/10.1103/PhysRevX.11.041045} {\bibfield  {journal} {\bibinfo  {journal} {Phys. Rev. X}\ }\textbf {\bibinfo {volume} {11}},\ \bibinfo {pages} {041045} (\bibinfo {year} {2021})}\BibitemShut {NoStop}%
\bibitem [{\citenamefont {Marciniak}\ \emph {et~al.}(2022)\citenamefont {Marciniak}, \citenamefont {Feldker}, \citenamefont {Pogorelov}, \citenamefont {Kaubruegger}, \citenamefont {Vasilyev}, \citenamefont {van Bijnen}, \citenamefont {Schindler}, \citenamefont {Zoller}, \citenamefont {Blatt},\ and\ \citenamefont {Monz}}]{marciniak2022optimal}%
  \BibitemOpen
  \bibfield  {author} {\bibinfo {author} {\bibfnamefont {C.~D.}\ \bibnamefont {Marciniak}}, \bibinfo {author} {\bibfnamefont {T.}~\bibnamefont {Feldker}}, \bibinfo {author} {\bibfnamefont {I.}~\bibnamefont {Pogorelov}}, \bibinfo {author} {\bibfnamefont {R.}~\bibnamefont {Kaubruegger}}, \bibinfo {author} {\bibfnamefont {D.~V.}\ \bibnamefont {Vasilyev}}, \bibinfo {author} {\bibfnamefont {R.}~\bibnamefont {van Bijnen}}, \bibinfo {author} {\bibfnamefont {P.}~\bibnamefont {Schindler}}, \bibinfo {author} {\bibfnamefont {P.}~\bibnamefont {Zoller}}, \bibinfo {author} {\bibfnamefont {R.}~\bibnamefont {Blatt}},\ and\ \bibinfo {author} {\bibfnamefont {T.}~\bibnamefont {Monz}},\ }\bibfield  {title} {\bibinfo {title} {Optimal metrology with programmable quantum sensors},\ }\href {https://doi.org/10.1038/s41586-022-04435-4} {\bibfield  {journal} {\bibinfo  {journal} {Nature}\ }\textbf {\bibinfo {volume} {603}},\ \bibinfo {pages} {604} (\bibinfo {year} {2022})}\BibitemShut {NoStop}%
\bibitem [{\citenamefont {Sundar}\ \emph {et~al.}(2024)\citenamefont {Sundar}, \citenamefont {Barberena}, \citenamefont {Rey},\ and\ \citenamefont {Orioli}}]{sundar2024driven}%
  \BibitemOpen
  \bibfield  {author} {\bibinfo {author} {\bibfnamefont {B.}~\bibnamefont {Sundar}}, \bibinfo {author} {\bibfnamefont {D.}~\bibnamefont {Barberena}}, \bibinfo {author} {\bibfnamefont {A.~M.}\ \bibnamefont {Rey}},\ and\ \bibinfo {author} {\bibfnamefont {A.~P.~n.}\ \bibnamefont {Orioli}},\ }\bibfield  {title} {\bibinfo {title} {Driven-dissipative four-mode squeezing of multilevel atoms in an optical cavity},\ }\href {https://doi.org/10.1103/PhysRevA.109.013713} {\bibfield  {journal} {\bibinfo  {journal} {Phys. Rev. A}\ }\textbf {\bibinfo {volume} {109}},\ \bibinfo {pages} {013713} (\bibinfo {year} {2024})}\BibitemShut {NoStop}%
\end{thebibliography}
%

\end{document}